\documentclass[fleqn,usenatbib]{mnras}

\usepackage{newtxtext,newtxmath}

\usepackage[T1]{fontenc}

\DeclareRobustCommand{\VAN}[3]{#2}
\let\VANthebibliography\thebibliography
\def\thebibliography{\DeclareRobustCommand{\VAN}[3]{##3}\VANthebibliography}

\usepackage{graphicx}	
\usepackage{amsmath}	

\outer\def\gtae {$\buildrel {\lower3pt\hbox{$>$}} \over 
{\lower2pt\hbox{$\sim$}} $}
\outer\def\ltae {$\buildrel {\lower3pt\hbox{$<$}} \over 
{\lower2pt\hbox{$\sim$}} $}

\newcommand{\Rsun}{$R_{\odot}$}
\newcommand{\Rearth}{$R_{\oplus}$}

\newcommand{\tess}{\it TESS}

\newcommand{\rpl}{\mbox{R\textsubscript{p}}}
\newcommand{\rstar}{\mbox{R\textsubscript{*}}}

\title[The {\tess} SPOC FFI Sample]{The TESS SPOC FFI Target Sample Explored with Gaia \thanks{A copy of the full main sequence {\tess}-SPOC FFI target sample (see, Table \ref{tab:SPOC_ms_sample}) is available at CDS.  }}

\author[L. Doyle et al.]{
Lauren Doyle,$^{1,2}$\thanks{E-mail: lauren.doyle@warwick.ac.uk}
David J. Armstrong,$^{1,2}$
Daniel Bayliss,$^{1,2}$
Toby Rodel$^{1,2,3}$
and Vedad Kunovac$^{1,2}$
\\
$^{1}$Centre for Exoplanets and Habitability, University of Warwick, Coventry, CV4 7AL, UK \\
$^{2}$Department of Physics, University of Warwick, Coventry, CV4 7AL, UK \\
$^{3}$Astrophysics Research Centre, School of Mathematics and Physics, Queen’s University Belfast, Belfast, BT7 1NN, UK
}

\date{Accepted 2024 February 19. Received 2024 February 16; in original form 2023 October 11}

\pubyear{2024}

\begin{document}
\label{firstpage}
\pagerange{\pageref{firstpage}--\pageref{lastpage}}
\maketitle

\begin{abstract}
The {\tess} mission has provided the community with high-precision times series photometry for $\sim$2.8 million stars across the entire sky via the Full Frame Image (FFI) light curves produced by the {\tess} Science Processing Operations Centre (SPOC). This set of light curves is an extremely valuable resource for the discovery of transiting exoplanets and other stellar science. However, due to the sample selection, this set of light curves does not constitute a magnitude limited sample. In order to understand the effects of this sample selection, we use Gaia DR2 and DR3 to study the properties of the stars in the {\tess}-SPOC FFI light curve set, with the aim of providing vital context for further research using the sample. We report on the properties of the {\tess}-SPOC FFI Targets in Sectors 1 -- 55 (covering Cycles 1 -- 4). We cross-match the {\tess}-SPOC FFI Targets with the Gaia DR2 and DR3 catalogues of all targets brighter than Gaia magnitude 14 to understand the effects of sample selection on the overall stellar properties. This includes Gaia magnitude, parallax, radius, temperature, non-single star flags, luminosity, radial velocity and stellar surface gravity. In total, there are $\sim$16.7 million Gaia targets brighter than G=14, which when cross-matched with the {\tess}-SPOC FFI Targets leaves $\sim$2.75 million. We investigate the binarity of each {\tess}-SPOC FFI Target and calculate the radius detection limit from two detected {\tess} transits which could be detected around each target. Finally, we create a comprehensive main sequence {\tess}-SPOC FFI Target sample which can be utilised in future studies.
\end{abstract}




\section{Introduction}
Since its launch in April of 2018, the Transiting Exoplanet Survey Satellite \citep[{\tess}:][]{ricker2015tess} has transformed the field of stellar and exoplanetary physics. {\tess} has successfully discovered scores of transiting exoplanets, including planets transiting very bright host stars \citep[e.g.][]{gandolfi2018tess, huang2018tess}, small radius planets \citep[e.g.][]{gilbert2020first, oddo2023characterization}, multi-planet systems \citep[e.g.][]{leleu2021six} and planets around young stars \citep[e.g.][]{newton2019tess, battley2020search, mann2022tess}. At the time of writing, {\tess} is currently in its second extended mission, observing Sectors 70 -83. This marks the sixth year of {\tess} observations observing some of the ecliptic and the Northern hemisphere for a third time \footnote{\href{https://tess.mit.edu/tess-year-6-observations/}{https://tess.mit.edu/tess-year-6-observations/}}. The extended {\tess} mission allows for the detection of longer-period transiting planets \citep[e.g.][]{gill2020ngts, lendl2020toi} and also opens the door into the search for stellar activity cycles \cite[e.g.][]{davenport202010, doyle2022puzzling}. 

{\tess} is composed of four cameras, each of which maps on to an array of four CCDs. The combined field-of-view (FOV) is 24$^{\circ} \times$ 94$^{\circ}$ which makes up one Sector where each is observed for $\sim$27 days. For a full description of the {\tess} mission, see \citet{ricker2015tess}. There are various data collection modes for {\tess}, however, for this work we exclusively focus on the Full Frame Images (FFIs). For Sectors 1-26 the FFIs were acquired every 30\,min where for Sectors 27-56 the cadence of the FFIs was improved to every 10\,min. From Sector 57 onwards, the cadence was improved even further to 200\,s, making this mode comparable to the 2-minute postage stamp targeted sample. The {\tess} Science Processing Operations Centre \citep[SPOC:][]{jenkins2016tessspoc} is responsible for the {\tess} Science pipeline which is based off the very successful Kepler pipeline. From the second year of the {\tess} mission, the SPOC began processing FFI light curves for up to 160,000 targets per Sector \citep{caldwell2020spoctargets}. All pixel and light curve data for the {\tess}-SPOC FFI Target sample are contributed as High-Level Science Products (HLSP) on the Mikulski Archive for Space Telescopes (MAST\footnote{\href{https://archive.stsci.edu/}{https://archive.stsci.edu/}}) where they are publicly available. Due to the fact that the {\tess}-SPOC FFI constitute a large and homogeneously processed set of light curves, they have been an extremely valuable resource for the community, used for a wide range of studies including planet discoveries \citep[e.g.][]{gilbert2020first, eisner2021planet, yee2022tess}, statistical studies \citep[e.g.][]{bryant2023occurrence}, and binary star studies \citep[e.g.][]{ijspeert2021all, prvsa2022tess}.

In this paper, we explore the physical parameters of the stars in the {\tess}-SPOC FFI Target sample using observations from the Gaia mission \citep{prusti2016gaia}, both from Gaia DR2 \citep{brown2018gaiadr2} and DR3 \citep{vallenari2023gaiadr3}. In \S \ref{sec:target_samples} we detail the full Gaia and SPOC samples and discuss how we cross-match them to achieve our final {\tess}-SPOC FFI Target sample. Here, we also look at the spread of observed {\tess} Sectors for each target. In \S \ref{sec:planet_detections} we determine the radius detection limit from two detected {\tess} transits which could be detected around each {\tess}-SPOC FFI target and also look at the distribution of {\tess} Objects of Interest (TOIs) in the colour-magnitude diagram of the {\tess}-SPOC FFI Target sample. For \S \ref{sec:binary} we investigate the binarity of the sample using the non-single star and RUWE flags in the Gaia data. Finally in \S \ref{sec:SPOC_ms} we isolate main sequence stars in the {\tess}-SPOC FFI Target sample as a prime sample for further future statistical studies that wish to focus on dwarf stars. 

\section{Target Samples}
\label{sec:target_samples}
The precise method of sample selection for stars to be included in the {\tess}-SPOC FFI light curves results in a set of targets that is close to, but not quite, a magnitude limited sample. In this Section, we detail both the {\tess}-SPOC FFI Target sample, a Gaia magnitude limited sample, and investigate the difference between the two. 

\subsection{{\tess}-SPOC FFI Sample}
\label{sec:SPOC_sample}

The {\tess}-SPOC FFI targets are selected on a set of criteria designed to maximise scientific goals and minimise the impact on processing times in the SPOC context \citep{caldwell2020spoctargets}. The target selection is made in the following way:

\begin{enumerate}
    \item  All two-minute targets are selected ($\sim$20,000 per Sector).
    \item Targets are selected with a H magnitude $\leq$10 or with a distance $\leq$100~pc, provided the crowding metric $\geq$0.5 and the TESS magnitude $\leq$16. 
    \item Targets with TESS magnitude $\leq$13.5, log surface gravity ($\log{g}$) $\leq$3.5, and crowding metric $\leq$0.8.
\end{enumerate}     

The crowding metric, as defined in the SPOC and Kepler pipelines \citep{smith2016finding}, the fraction of flux in the optimal aperture that is due to the target star. For example, a crowding metric of 0.8 means 80\% of the flux within the aperture is from the target star, therefore, in (ii) and (iii) above the crowding metric must be greater than 0.5 and 0.8 respectively for the target to be selected.

\begin{figure}
    \centering
    \includegraphics[width = 0.47\textwidth]{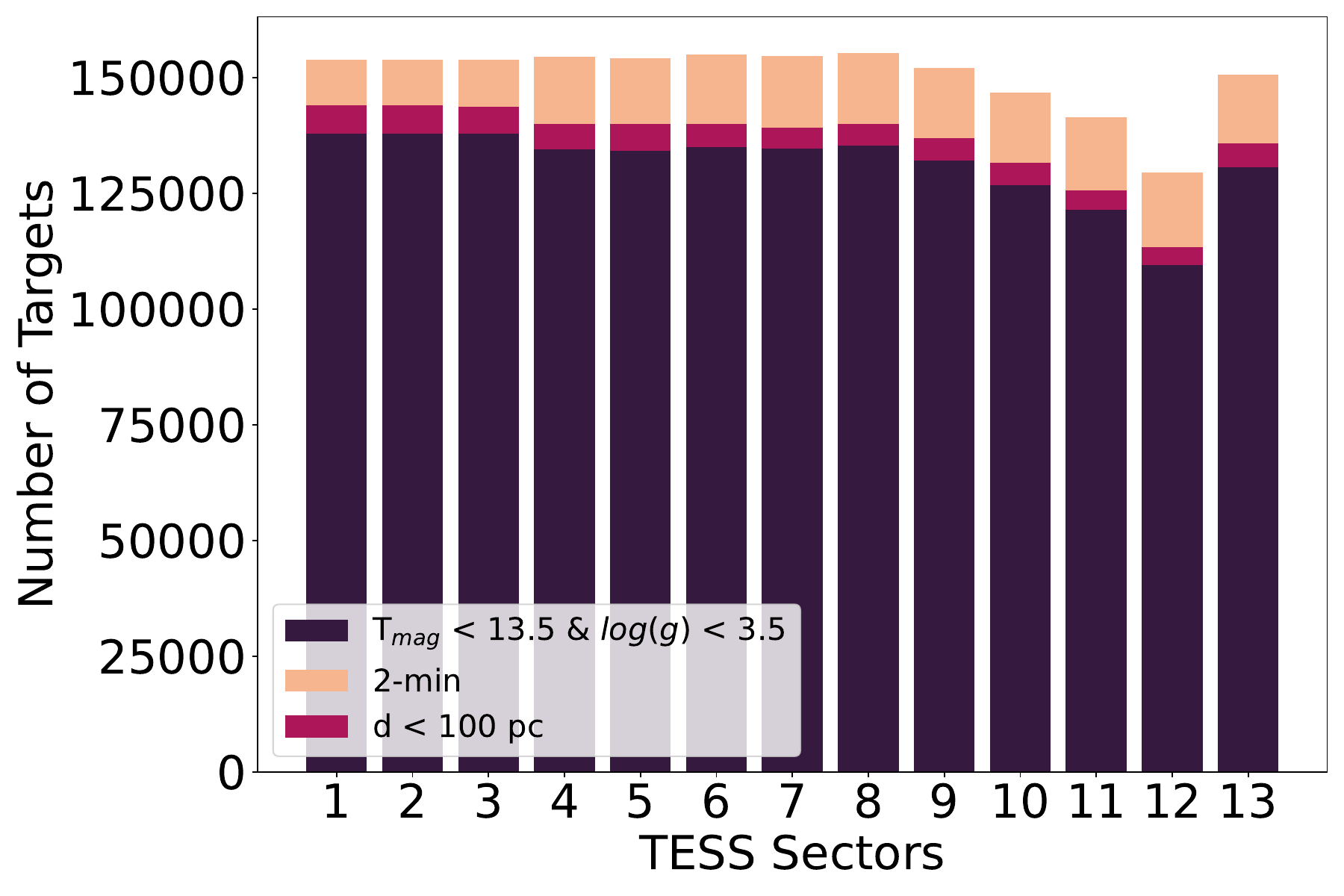}
    \caption{A bar chart showing the split of {\tess}-SPOC FFI Targets in each sector for Cycle 1. In orange are the targets which are also observed in 2-min cadence (10 -- 15\%), pink are the targets with d $\leq$100~pc (3 -- 4\%) and in purple the remaining targets which represent those with T$_{\rm{mag}}$ $\leq$13.5, $\log{g} \leq$3.5, and crowding metric $\leq$0.8. }
    \label{fig:targets_sector}
\end{figure}

Targets are selected on a per CCD basis, following the priority order set out above. A limit of 10,000 targets per CCD is placed on the allocation to ensure that the processing time does not impact the operations of the SPOC. Given {\tess} contains a total of 16 CCDs, this results in a limit of 160,000 stars for the {\tess}-SPOC FFI light curves for a given sector. Many times this limit is indeed reached, which in practice means that there will be stars in the 3rd select category above which are not included. In such cases, the targets are selected by apparent magnitude, with the brightest stars being prioritised. The split of {\tess}-SPOC FFI targets for each sector in Cycle 1 is shown in Figure \ref{fig:targets_sector}. From this it is clear the 2-min targets account for 10 -- 15\%, the M dwarf sample with d$\leq$100~pc accounts for 3--4\% and the vast majority are targets with TESS magnitude $\leq$13.5, log surface gravity $\leq$3.5, and crowding metric $\leq$0.8. It is also worth noting there is a dip in the overall number of {\tess}-SPOC FFI targets between sectors 8 and 12 which is a result of the galactic plane being withing the {\tess} field of view, resulting in less targets meeting the crowding metric cut of 0.8. 

\begin{figure}
    \centering
    \includegraphics[width = 0.47\textwidth]{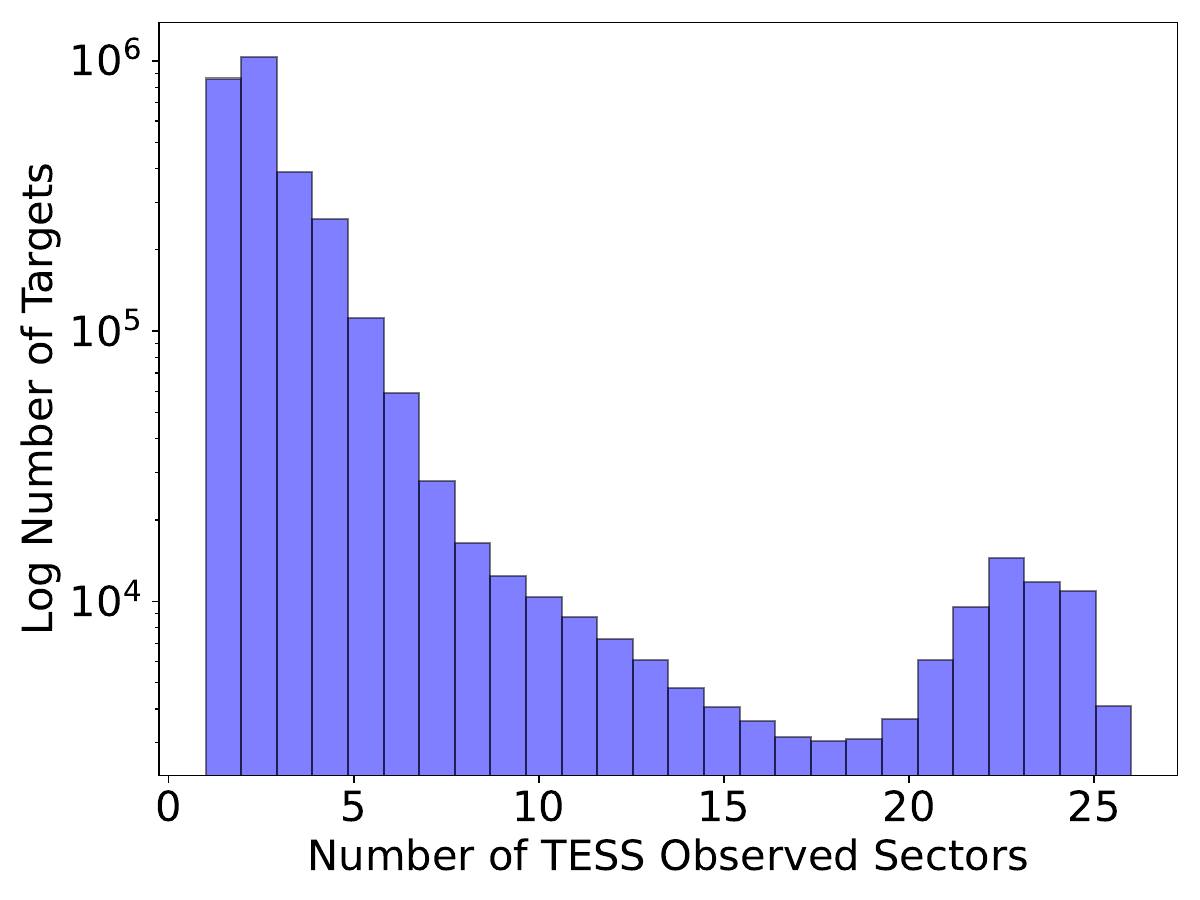}
    \caption{The histogram of the observed {\tess} Sectors of each {\tess}-SPOC FFI target in a logarithmic scale. This shows the spread of observed Sectors in the sample with the majority being observed between 2 and 3 Sectors. There is a further peak in the histogram at 23 Sectors which represents stars within the {\tess} continuous viewing zone. }
    \label{fig:TESS_sector_hist}
\end{figure}

In this study, we select all {\tess}-SPOC FFI targets which were observed in Sectors 1 -- 55, covering Cycles 1 -- 4. This is due to {\tess} currently observing in Cycle 5/6 where all data is not yet available at the time of writing. Furthermore, from Cycle 5 onwards {\tess} will observe FFIs in 200-second cadence. By using observations from Cycles 1 -- 4, we have observations covering both northern and southern hemispheres twice, creating a complete sample which will allow for repeatability of findings between sectors and cycles. 

For all of the {\tess}-SPOC FFI targets in Sectors 1 -- 55, we gather their TIC ID, right ascension, declination and their Gaia DR2 identifier which compromises a total of 2,891,782 individual targets. We also determine the number of {\tess} Sectors each individual target has been observed for and show the results in Figure \ref{fig:TESS_sector_hist}. It can be seen that the majority of targets are observed for 2 to 3 Sectors where there is a peak in the histogram. There is a further peak at around 23 Sectors which represents targets within the ecliptic poles that lie within the {\tess} continuous viewing zone. Lastly, we also extract the precision of each target from the header of the SPOC light curve and use this information in \S \ref{sec:planet_detections}.

\subsection{Gaia Sample}
\label{sec:Gaia_sample}

The Gaia mission \citep{prusti2016gaia} is part of the science programme of the European Space Agency (ESA) and was launched in December of 2013. The scientific goals of Gaia are extensive where the key aim is to measure the distances, positions, space motions and physical properties of one billion stars in our galaxy. Gaia provides a map of our galaxy for the first time, complete for all stars brighter than G=20. This provides detailed information on all stars in the {\tess}-SPOC FFI Target sample set out in Section~\ref{sec:SPOC_sample}. For this study, we use data released by Gaia in DR2 \citep{brown2018gaiadr2} and DR3 \citep{vallenari2023gaiadr3} accessing the data through a Table Access Protocol (TAP) query using {\tt ADQL} in {\tt TOPCAT} \citep{taylor2005topcat}. 

Gaia DR2 is based on data collected during the first 22 months of the Gaia mission. This release made the leap to high precision parallax and proper motion measurements for over one billion stars, and is a major advancement on DR1 with regards to the completeness and accuracy. Homogeneous multi-band photometry and large-scale radial velocity observations at the bright (G $\le$ 13) end were also made available in the DR2 release. Gaia DR3 is based on EDR3 \citep[an earlier release in 2021:][]{brown2021gaia} where astrometry and broad band photometry were not updated from DR2. The full Gaia DR3 was released in 2022 with updated radial velocities, spectra from the radial velocity spectrograph, astrophysical parameters for sources based on the blue and red prism photometer (BP and RP) spectra, amongst much more. This release was based on data gathered over the first 34 months of the mission and provides significant improvement in both the precision and accuracy of measurements. At the time of writing Gaia DR3 is the most recent Gaia data release, and the next Gaia data release (DR4) will not be released before the end of 2025\footnote{\url{https://www.cosmos.esa.int/web/gaia/release}}.

We use the Gaia DR3 parallaxes to estimate the absolute Gaia magnitude in the $G$ band for individual stars following the methods of \citet{babusiaux2018gaia}. In addition to retrieving the DR3 sample we also fetched the DR2 identifiers (\texttt{source\_id}) which is very useful in cross-matching the {\tess}-SPOC FFI sample. This is because the {\tess} Input Catalogue \citep[TIC v8.2:][]{stassun2019tic} IDs are associated with the Gaia DR2 identifiers. Between Gaia DR2 and DR3 there are changes to the identifiers just as there are between Gaia DR1 and DR2. This is as a result of the source lists becoming progressively more stable and of higher spatial resolution over the course of the Gaia mission. 

The stellar astrophysical parameters contained within the Gaia DR3 release are comprised of atmospheric properties, evolutionary parameters, metallicity, individual chemical element abundances, and extinction parameters, along with other characterisation such as equivalent widths of the H$\alpha$ line and activity index for cool active stars \citep[see][for full details]{vallenari2023gaiadr3}. In Gaia DR2, median radial velocities for $\sim$ 7 million sources were presented, along with estimates of the stellar effective temperature, extinction, reddening, radius and luminosity for between 77 and 161 million sources \citep[see][for full details]{brown2018gaiadr2}. Gaia DR3 contained newly determined radial velocities for about 33.8 million stars with $G_{\rm{RVS}}$ (median of the single-transit radial velocity spectrometer measurements) $\le$ 14 and with 3100 K $\le$ $T_{\rm{eff}}$ $\le$ 14,500~K. Therefore, we use the radial velocity, parallax and $\log{g}$ measurements from DR3 and the effective temperature, stellar radius and luminosity from DR2. 

The Gaia DR2 effective temperatures are estimated from two distance-independent colours $G_{\rm{BP}} - G$ and $G - G_{\rm{RP}}$. Machine learning is used to come up with a non-parametric model for the colour-temperature relation where the training sample contains targets with $T_{\rm{eff}} = $3,000 -- 10,000~K. Therefore, there is no Gaia target with a temperature listed outside of this range. The luminosities from Gaia DR2 are estimated using the \texttt{FLAME} software package \citep[part of the Apsis data processing system, see][for further details]{bailer2013gaia} utilising relations between the effective temperature and Gaia absolute magnitude. Finally, the stellar radius is then estimated using the effective temperature and luminosity. 

\begin{figure}
    \centering
    \includegraphics[width = 0.47\textwidth]{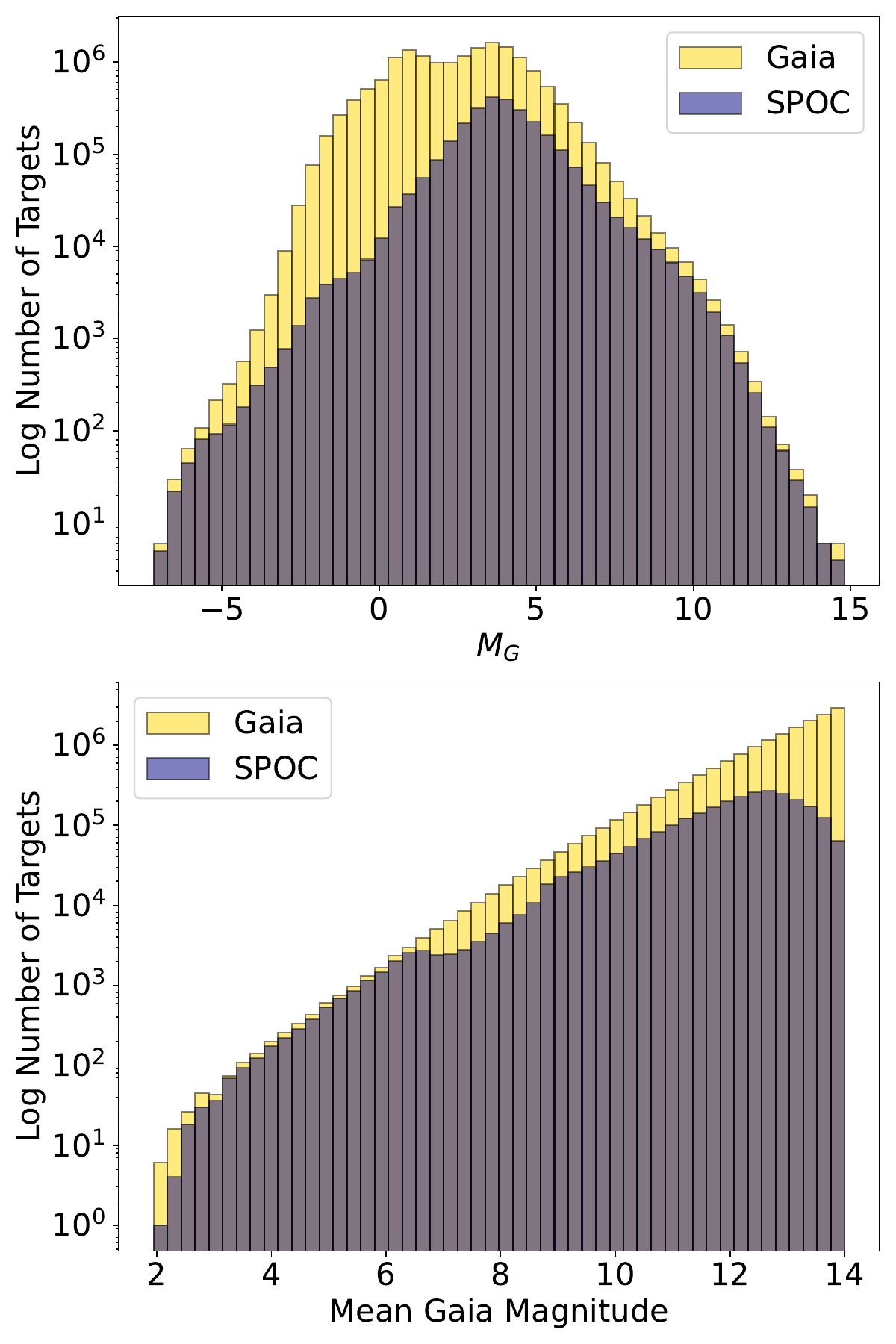}
    \caption{A histogram of the absolute Gaia magnitude and mean Gaia magnitude from DR3 for both the Gaia sample (yellow) and the {\tess}-SPOC FFI Target sample (navy).}
    \label{fig:Gaia_hist}
\end{figure}

\begin{figure}
    \centering
    \includegraphics[width = 0.47\textwidth]{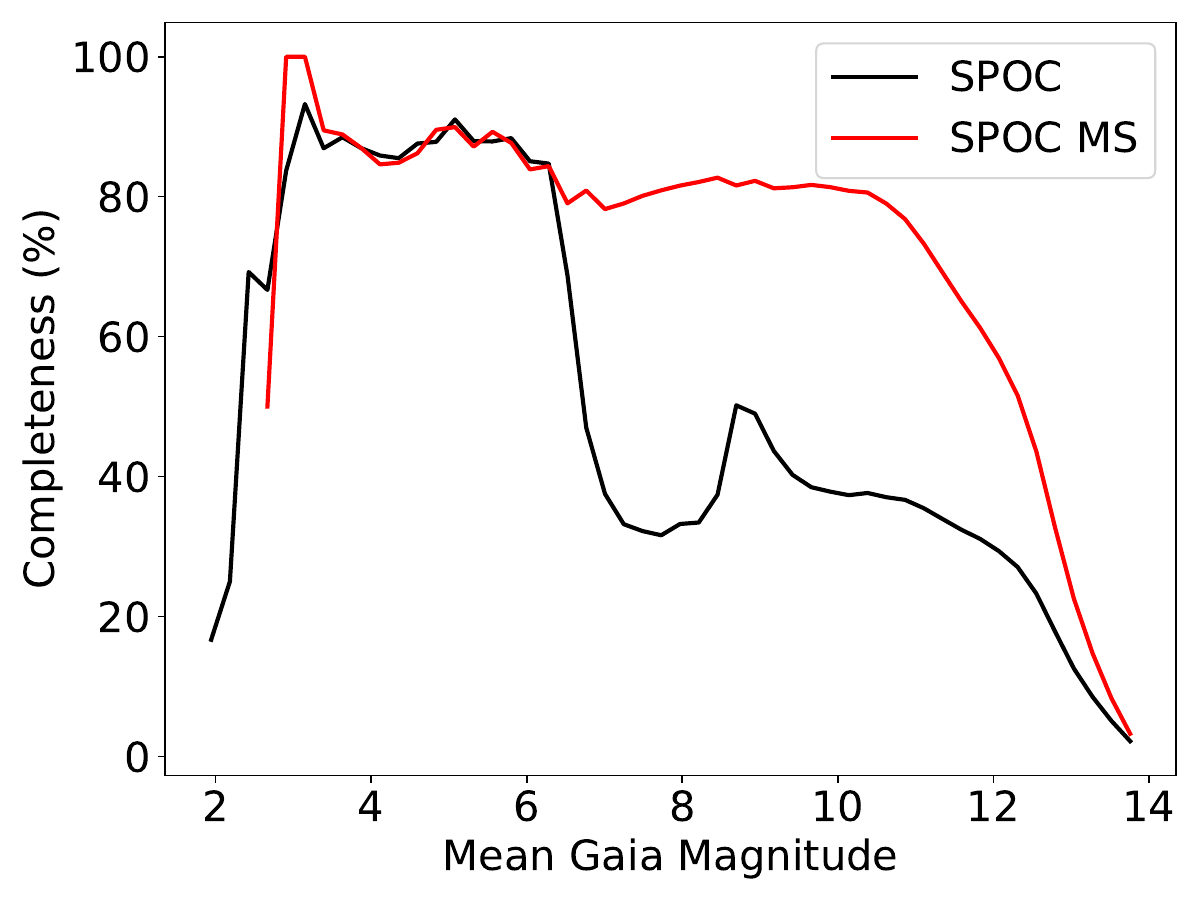}
    \caption{The completeness of the {\tess}-SPOC FFI Target sample in comparison to the Gaia magnitude limited sample. The black line represents the full {\tess}-SPOC FFI sample compared to the full Gaia sample and the red line is the {\tess}-SPOC FFI main sequence sample compared to the Gaia main sequence sample defined in \S \ref{sec:SPOC_ms}. }
    \label{fig:completeness}
\end{figure}

\begin{figure*}
    \centering
    \includegraphics[width = 0.97\textwidth]{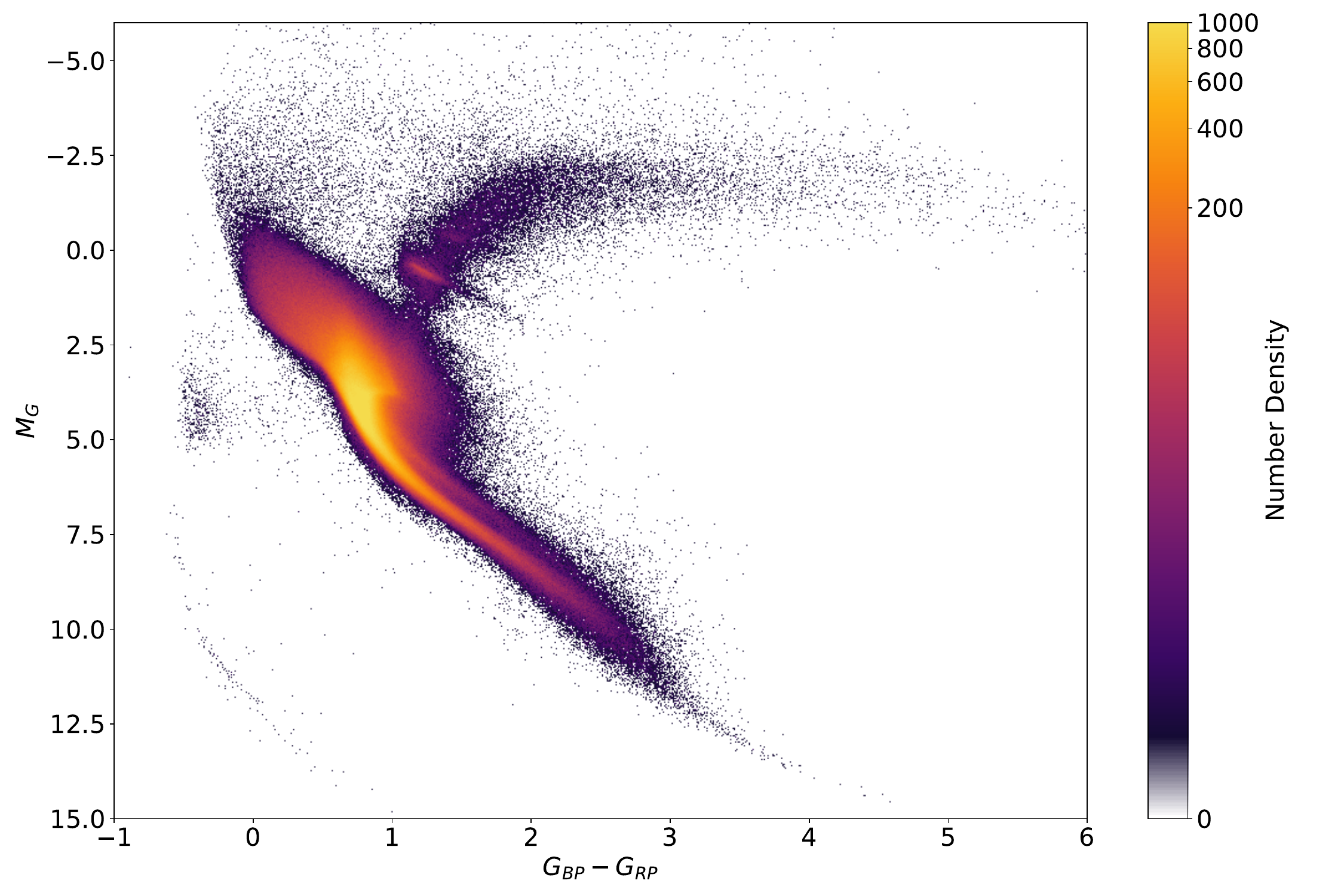}
    \caption{The colour-magnitude diagram of all {\tess}-SPOC FFI Targets from {\tess} Sectors 1 -- 55. All colours along with the parallax, used to determine the absolute Gaia magnitude, were taken from the Gaia DR3 catalogue. The colour scale represents the log of the density of stars. }
    \label{fig:SPOC_HRD}
\end{figure*}

\subsection{{\tess}-SPOC FFI and Gaia Cross-Matched Sample}
\label{sec:final_sample}
In order to cross-match the {\tess}-SPOC FFI Target sample with Gaia, we first create a magnitude limited sample of Gaia stars in DR3 where the relative precision on the parallax is better than 20\% (following \citet{babusiaux2018gaia}) and $G<$14. This brightness cut is used as the {\tess}-SPOC FFI Target sample predominantly comprises stars brighter than $T_{\rm{mag}} = \sim$ 13.5 (see \S \ref{sec:SPOC_sample} for further details). Furthermore, the number of Gaia targets fainter than G=14 increases significantly, making it more difficult and time consuming to access and download the Gaia sample. We cross-matched the {\tess}-SPOC FFI Targets to the Gaia magnitude limited sample using the Gaia DR2 source identifiers. This produced a total of 2,744,013 cross-matches, leaving 147,769 {\tess}-SPOC FFI targets without a match. This deficit results from the two cuts on magnitude (94.7\%) and parallax error precision (5.3\%) made when creating the Gaia magnitude limited sample. In order to form a more complete sample of the {\tess}-SPOC FFI targets we cross matched the remaining 147,769 with the Gaia DR3 catalogue directly using their right ascension and declination from the TIC and a radius search of 8 arcsec. This produced a further 146,604 {\tess}-SPOC FFI targets with Gaia DR3 parameters where the 1,165 missing targets were a result of high proper motion on the order of hundreds of milliarcseconds per year. There were a number of targets which were flagged as duplicates by {\tess} and these have been removed. The original cross match and the additional targets from the radius search make the full final {\tess}-SPOC FFI Target sample with a total count of 2,890,583 individual targets. 

\begin{figure}
    \centering
    \includegraphics[width = 0.47\textwidth]{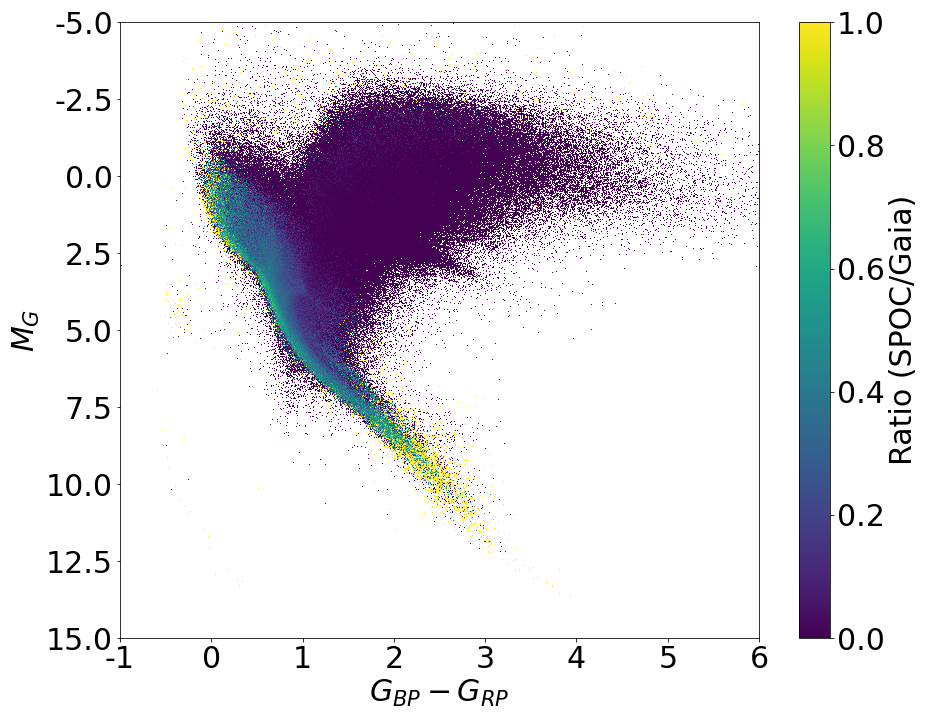}
    \caption{Colour-magnitude diagram where the colour scale shows the ratio between {\tess}-SPOC FFI and Gaia targets. Yellow represents both samples having the same number of targets and dark blue where targets are in the Gaia sample but not the {\tess}-SPOC FFI sample.}
    \label{fig:SPOC_Gaia_comp}
\end{figure}

\begin{figure*}
    \centering
    \includegraphics[width = 0.97\textwidth]{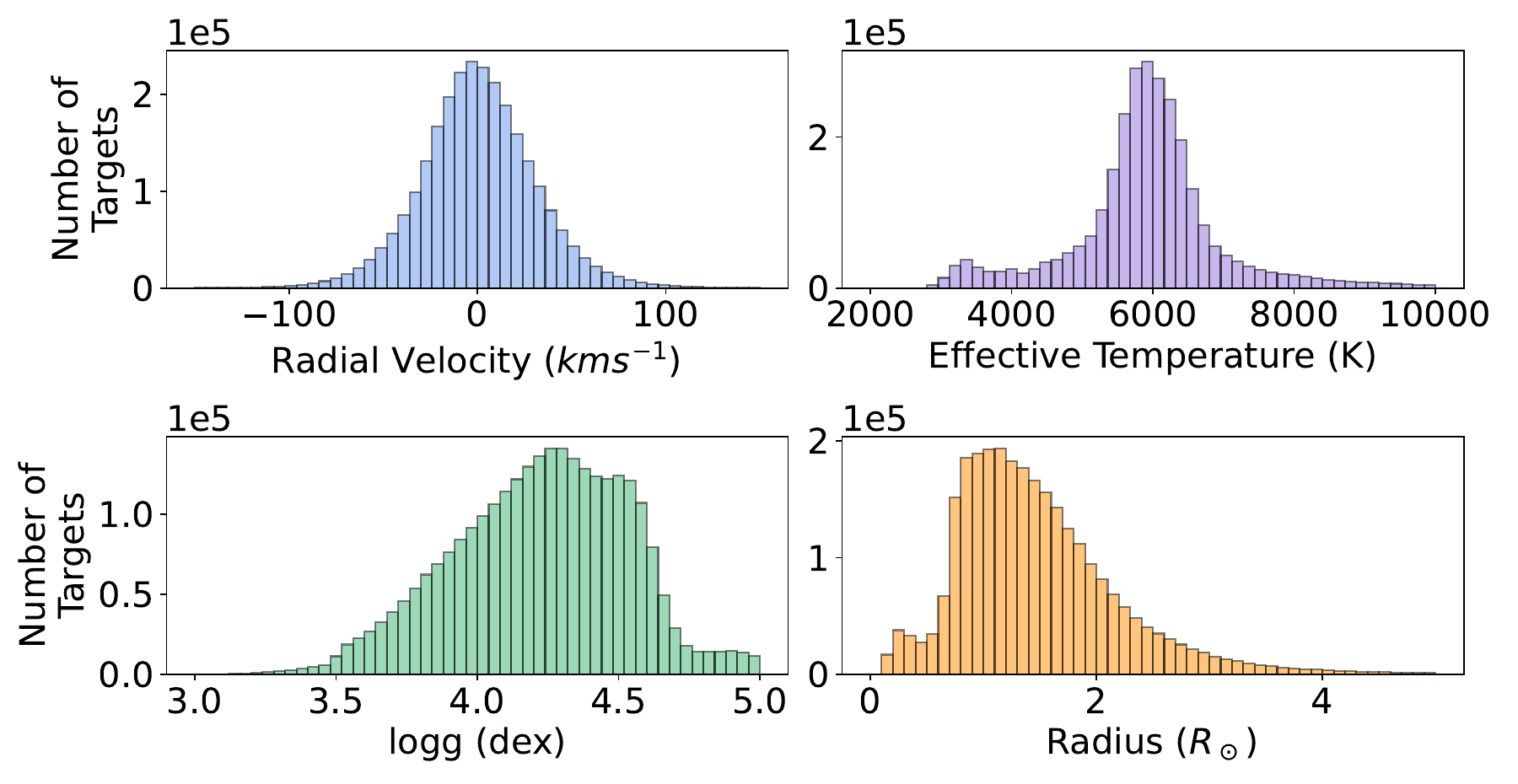}
    \caption{Stellar properties of the {\tess}-SPOC FFI Target sample. The radial velocity measurements are from Gaia DR3 and the remaining properties are taken from the Gaia DR2 catalogue. The range shown in each plot was cut to enable the distribution to be seen. The percentage of the sample which lies out of the axes is as follows; radial velocity 3.7\%, temperature 1.8\%, $\log{g}$ 2.5\% and radius 1.9\%.}
    \label{fig:SPOC_prop_hist}
\end{figure*}

\begin{figure}
    \centering
    \includegraphics[width = 0.47\textwidth]{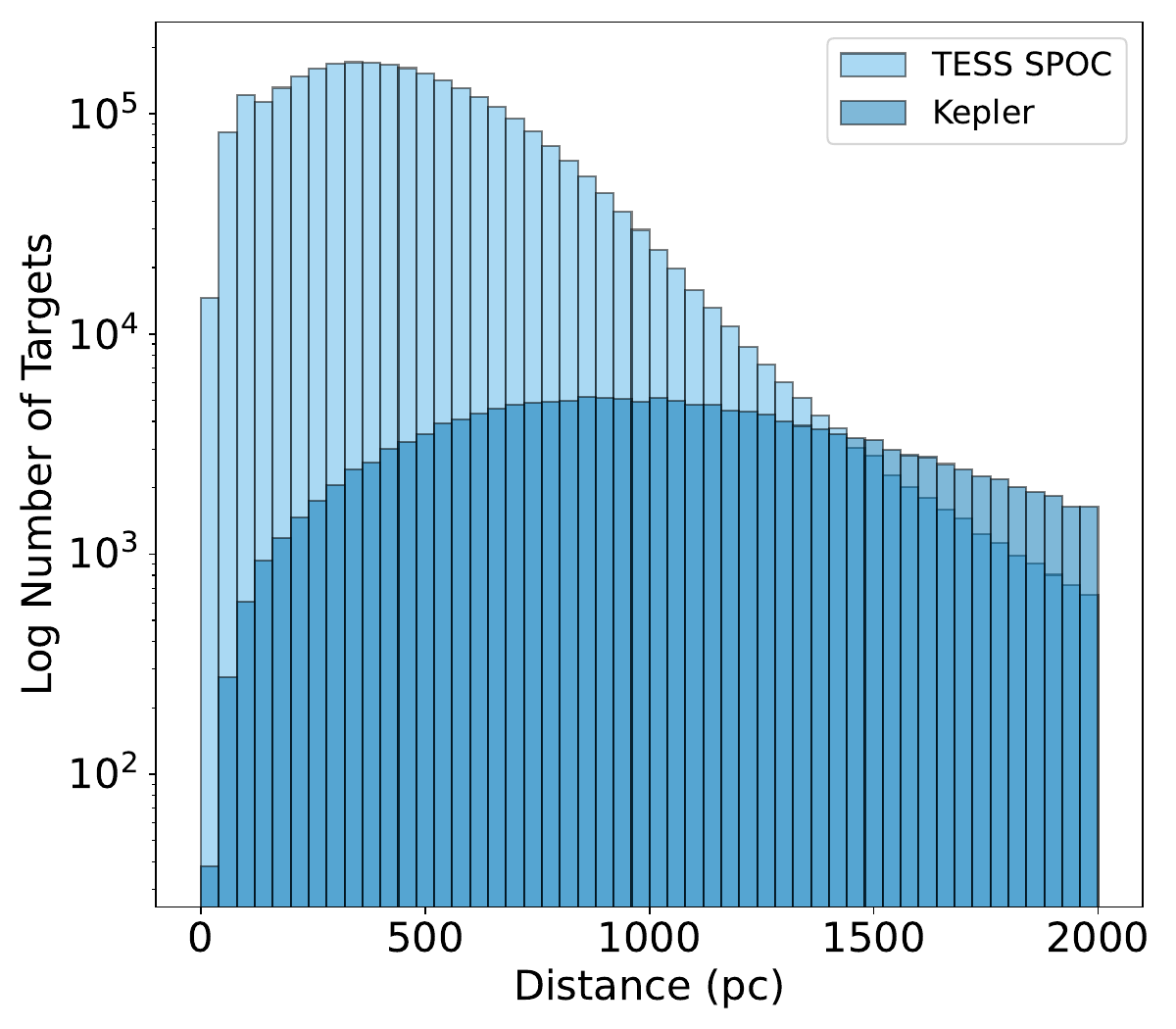}
    \caption{The log distribution of distances for the {\tess}-SPOC FFI Target sample, taken from the parallaxes of Gaia DR3 (light blue), showing those with distances less than 2,000~pc. We also overplot the {\sl Kepler} distances of $\sim$200,000 targets from Q1 (dark blue). The range shown has been cut to enable the distribution to be seen, therefore, 0.6\% of TESS and 22\% of {\sl Kepler} targets lie outside the axes. }
    \label{fig:SPOC_distances}
\end{figure}

Figure \ref{fig:Gaia_hist} shows the spread of absolute and mean Gaia magnitudes in the {\tess}-SPOC FFI Target sample compared with the underlying Gaia sample for G$<$14. All stars which fall into the {\tess} field of view and are brighter than T=6 (but not so bright as to cause CCD saturation) are selected for 2\,min cadence light curves \citep{stassun2018tess}. Therefore, since all of these 2\,min cadence targets are also chosen as {\tess}-SPOC FFI targets (see selection criteria set out in Section~\ref{sec:SPOC_sample}), the {\tess}-SPOC FFI Target sample is complete for almost all stars with G$<$6.
 
For stars fainter than G=6, the {\tess}-SPOC FFI Target sample is incomplete, and the degree of incompleteness is shown in Figure \ref{fig:completeness}. In this plot, the black line represents the percentage of completeness within the {\tess}-SPOC FFI Target sample compared to the Gaia magnitude limited sample. The red line then shows the completeness of the {\tess}-SPOC FFI main sequence sample (defined in \S \ref{sec:SPOC_ms}) compared to the Gaia magnitude limited main sequence sample which is created in the same way as the SPOC sample with a cut on surface gravity $\log{g} > $~3.5. From this, it is clear the full {\tess}-SPOC FFI Target sample is almost complete for stars with G$<$6 where it drops of suddenly. There is also a peak in completeness around G=9 which could correspond to large {\tess} Guest Investigator (GI) programmes as these targets are observed in 2\,min cadence and so are automatically included in the {\tess}-SPOC FFI Target sample. For the main sequence sample there is a much higher completeness across the whole magnitude range which drops off towards the end at G=13. 

\begin{figure*}
    \centering
    \includegraphics[width = 0.97\textwidth]{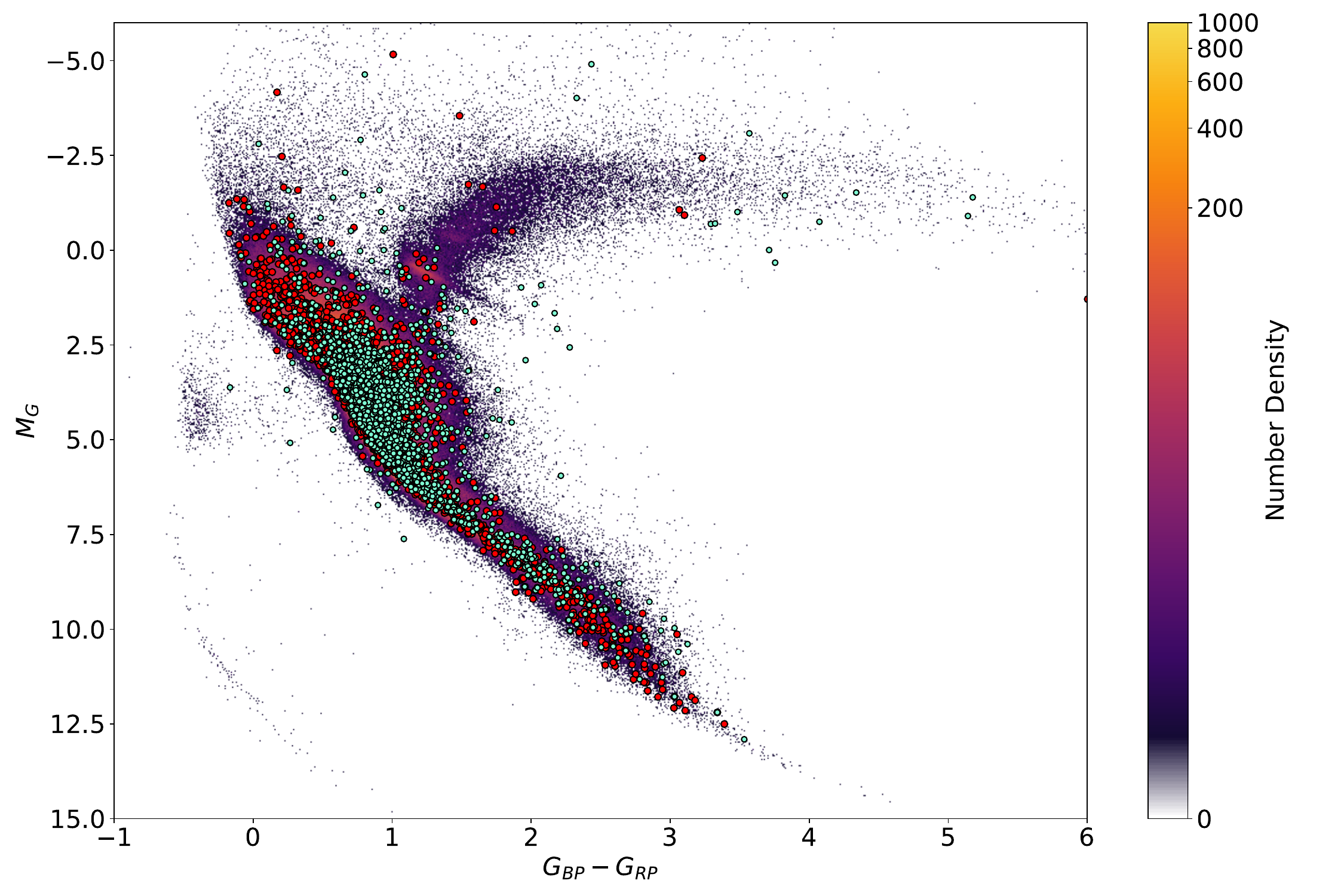}
    \caption{{\tess} Targets of Interest (TOIs) over plotted as circles onto the colour-magnitude diagram of all {\tess}-SPOC FFI Targets. Those with SPOC data are plotted in red (4,270) with the remaining TOIs with no SPOC data over-plotted in green (2,417). }
    \label{fig:TESS_TOIs}
\end{figure*}

Using the Gaia DR3 magnitudes, parallaxes and colours we are able to plot the full Hertzsprung Russell Diagram (HRD) for the {\tess}-SPOC FFI Target sample which can be seen in Figure \ref{fig:SPOC_HRD}. This shows the diversity of stellar classes and types which are present within the {\tess}-SPOC FFI Target sample. The majority of the sample are on the main sequence. At 0.75 magnitude above the main sequence, a subtle binary track can be seen which is due to near-equal mass main sequence binary stars. Also evident on the HRD are the spread of sub-giants and giants, a clump of blue hot-sub dwarfs and the faint white dwarf sequence. All of these classes of stars are present in the {\tess}-SPOC FFI Target sample due to being selected as 2-minute cadence targets. This selection is due to the stars being very bright (T$<$6) or through TESS GI programs that target specific classes of stars such as white dwarfs in G04137 and G022028 by JJ Hermes, subgiants in G022099 by J. Tayar and hot-sub dwarfs in G022141 by B. Barlow. 

Figure \ref{fig:SPOC_Gaia_comp} shows the ratio of {\tess}-SPOC FFI targets to the magnitude limited Gaia sample in the format of an HRD with a gird resolution of 2,000 $\times$ 2,000. Yellow colouring indicates the {\tess}-SPOC FFI Target sample has the same number of targets as the underlying Gaia sample - i.e. the {\tess}-SPOC FFI sample is approximately complete. Dark blue colouring indicates there are very few {\tess}-SPOC FFI targets compared with the underlying Gaia sample - i.e. the {\tess}-SPOC FFI sample is very incomplete. It is evident that the {\tess}-SPOC FFI Target sample is most complete for fainter main sequence low mass stars and also across the main sequence in general. The {\tess}-SPOC FFI Target sample is most incomplete for evolved/giant stars. This is a result of the TESS 2-minute targets being largely free of giant stars \cite[see discussions in ][on the assembly of the Candidate Target List]{stassun2018tess}. The reason for this was that {\tess} is an exoplanet transit survey mission, and transits around sub-giants or giants are much shallower compared to equivalent transits around main-sequence stars. Additionally, the criteria in which additional {\tess}-SPOC FFI targets are selected (see \S \ref{sec:SPOC_sample}) include nearby stars (d$<$100\,pc) and exclude low surface gravity stars ($\log g>3.5$), both of which will naturally exclude horizontal and giant branch stars. 

The distribution of Gaia stellar properties for the {\tess}-SPOC FFI Target sample are set out in Figure~\ref{fig:SPOC_prop_hist}. The distributions in radial velocity is a Gaussian with a peak at 0~kms$^{-1}$, in line with the expectation for a large set of nearby stars selected from across the entire sky. The distribution of effective temperatures is also sharply peaked at around 6,000\,K, reflecting the fact that the sample is largely made up of solar-type main sequence stars. There is a small bump in the distribution at approximately 3500\,K, which comes about both from the selection of M-dwarfs in the 2\,min frames and the selection of additional M-dwarfs in the {\tess}-SPOC FFI selection for nearby stars (d$<$100). Similarly the $\log g$ and stellar radius distributions largely reflect the fact that dwarf stars on the main sequence make up the bulk of the sample. Figure \ref{fig:SPOC_distances} shows the distributions of distances which peaks at around 400\,pc, with 90\% of targets having a distance of 850\,pc or less. Furthermore, we also plot the distribution of distances for $\sim$200,000 Kepler targets observed in Q1 where there is a peak at 1,000~pc. It is important to note here that {\tess} is observing many more stars at 1,000~pc compared to {\sl Kepler}, however, from 1,500~pc onwards {\sl Kepler} begins to take over observing more stars at a further distance. A small peak can be seen for $d<100$\,pc, which is due to the selection of nearby stars in 2-minute sample and in the {\tess}-SPOC FFI selection criteria.

\section{Exoplanet detections with {\tess} data}
\label{sec:planet_detections}

\begin{figure*}
    \centering
    \includegraphics[width = 0.97\textwidth]{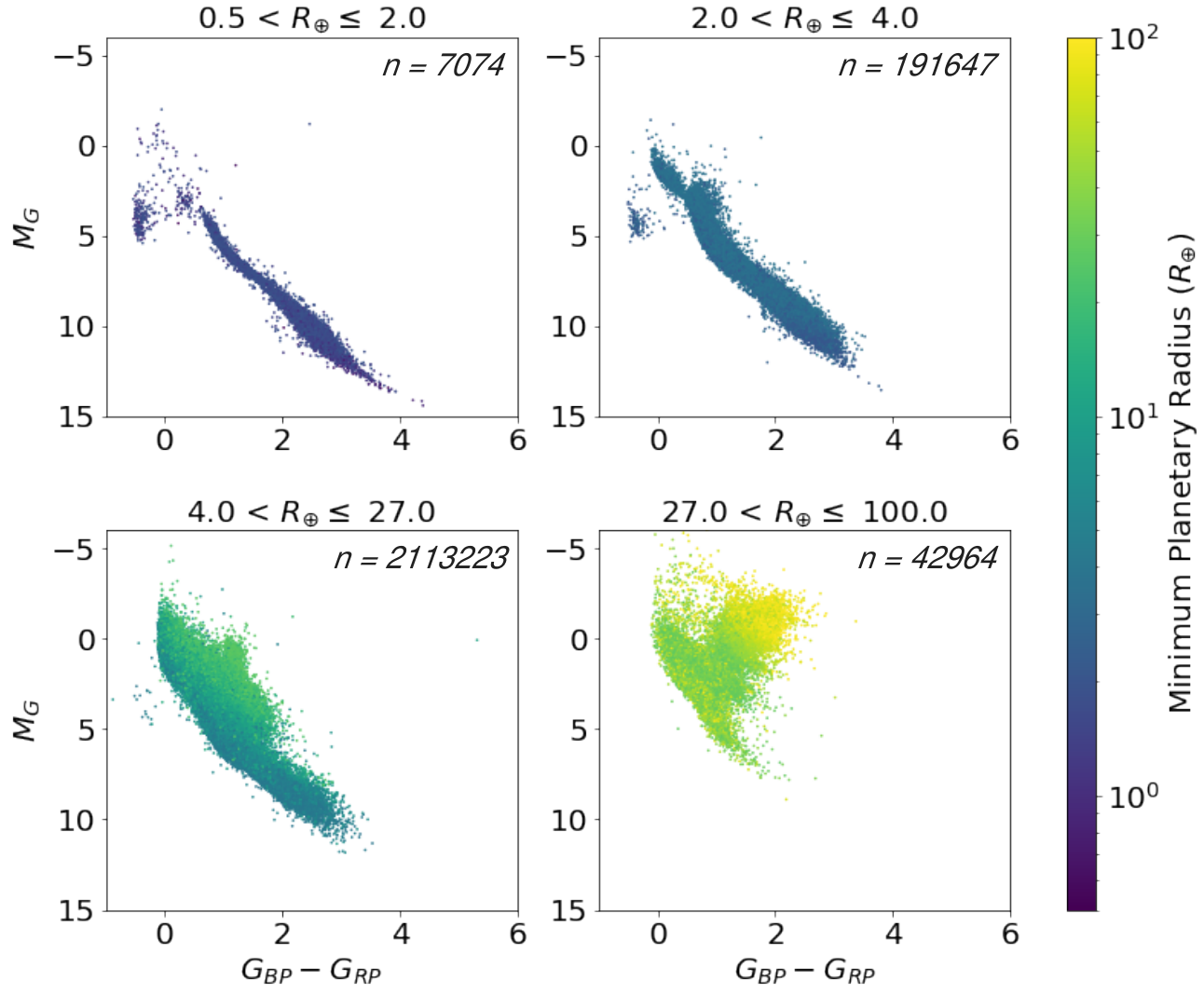}
    \caption{Colour-magnitude diagrams showing the distribution of the calculated two transit radius detection limit of an exoplanet which could be detected around each {\tess}-SPOC FFI target. Each plot has been split up according to planetary radius with Earths/Super-Earths in the range 0.5 $<$ \Rearth $\leq$ 2.0, mini-Neptunes in the range 2.0 $<$ \Rearth $\leq$ 4.0, Gas Giants in the range 4.0 $<$ \Rearth $\leq$ 27.0 and Non-Planetary Companions in the range 27.0 $<$ \Rearth $\leq$ 100.0. The number of each category in each subplot is given in the top right hand corner. The radius calculation was determined using the average TOI duration of 2.7~hrs as the transit duration, assuming a circular orbit and using the signal-to-noise of {\bf $7.1/\sqrt{2}$}.}
    \label{fig:planet_radius_HRD}
\end{figure*}

\begin{figure}
    \centering
    \includegraphics[width = 0.47\textwidth]{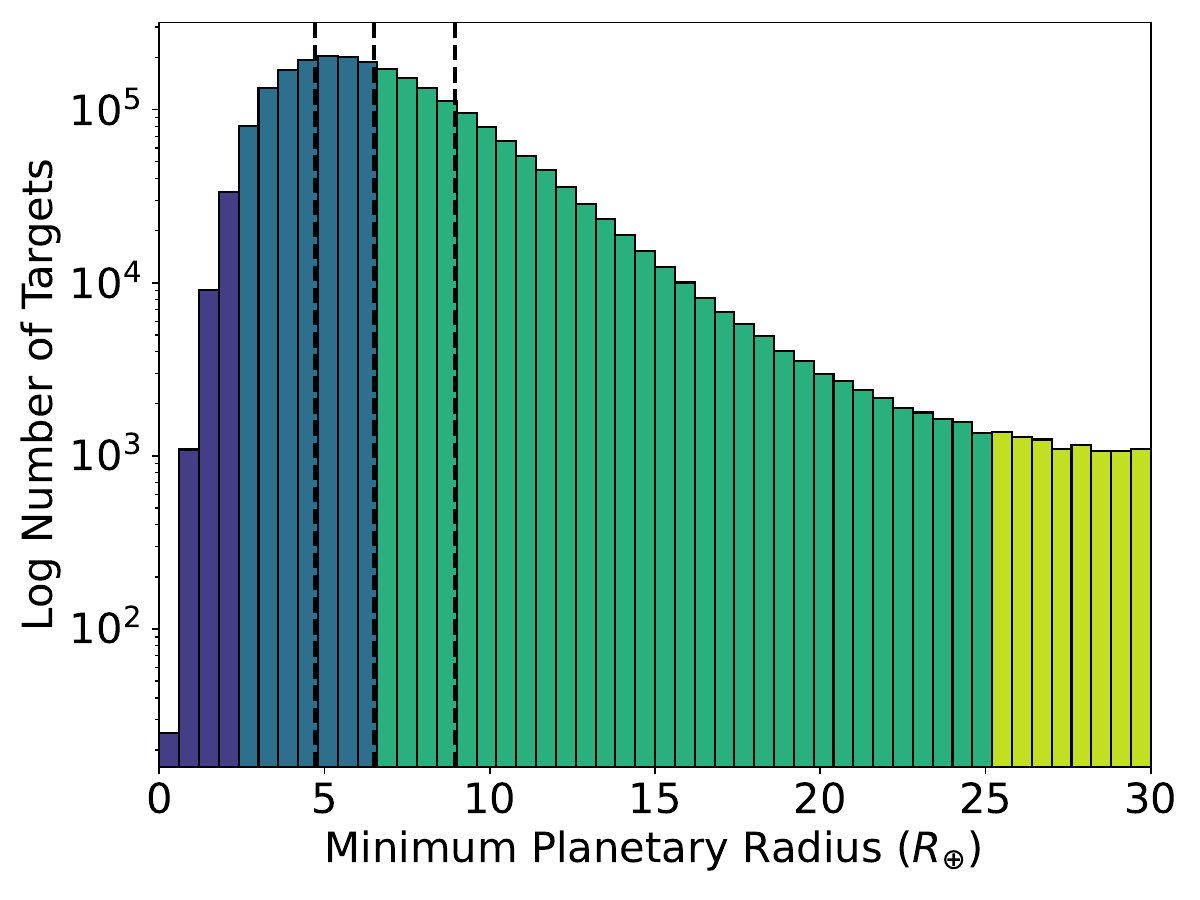}
    \caption{A log histogram distribution of the calculated two transit radius detection limit of an exoplanet which could be detected around each {\tess}-SPOC FFI Target. The radius calculation was determined using the median transit duration of all TOIs as the transit duration, assuming a circular orbit and using the signal-to-noise of $7.1/\sqrt{2}$. The bars are colour coded according to the planetary radius with Earths/Super-Earths in the range 0.5 $<$ \Rearth $\leq$ 2.0 as dark blue, Mini-Neptunes in the range 2.0 $<$ \Rearth $\leq$ 4.0 as blue/green, Gas Giants in the range 4.0 $<$ \Rearth $\leq$ 27.0 as green and Non-Planetary Companions in the range 27.0 $<$ \Rearth $\leq$ 100.0 as yellow. }
    \label{fig:SPOC_min_radius_hist}
\end{figure}

The primary science goal of {\tess} is to discover planets smaller than Neptune which transit stars bright enough to enable follow-up spectroscopic observations. It is this combination of photometry and spectroscopy which can provide properties of planetary systems such as planet radii, planet masses and atmospheric compositions. Therefore, the SPOC data products were developed with exoplanet transit searches as a focus.

\subsection{{\tess} Targets of Interest}
Potential transiting planets are found by searching for periodic flux decreases, known as Threshold Crossing Events (TCEs), in both the SPOC 2-min lightcurves created from postage stamps and 30-min cadence FFI light curves from the Quick Look Pipeline (QLP), which is another {\tess} data processing pipeline. These TCEs are then examined by the {\tess} Science Office (TSO) to identify planet candidates which would benefit from follow-up observations. Light curves are run through software to eliminate any obvious non-planetary signals where the remaining are manually vetted and listed as a TOI for follow up observations. Any TCEs which fall under other categories from eclipsing binaries to variable stars are not included in the TOI Catalog \citep{guerrero2021tois}, but are included in the comprehensive TCE Catalog available on MAST. At the time of writing there are currently 6,687 TOIs which have been followed-up and confirmed or are still awaiting further observations to confirm the planetary system. 

In \citet[Figure 6, ][]{guerrero2021tois} candidates from the {\tess} primary mission are over-plotted on a {\sl Gaia} sample. In a similar manner, we took the list of all currently known TOIs and cross matched it with the {\tess}-SPOC/Gaia FFI Target list. This produced a total of 3,950 known TOIs which have SPOC FFI data. For the remaining 2,737 targets we cross matched them directly with the Gaia DR3 catalogue to obtain the Gaia parameters for each. This was done using the right ascension and declination of each target and a cone search of 6 degrees. This then allowed us to plot the two TOI samples as those with SPOC FFI data (red) and those without (green) in Figure \ref{fig:TESS_TOIs} over top the SPOC HRD. While it is the 2-min TESS light curves that are used by TSO to identify TCE's, each 2-min light curve has a corresponding {\tess}-SPOC FFI light curve (see \S~\ref{sec:SPOC_sample}).  In Figure \ref{fig:TESS_TOIs} it can be seen that the majority of TOIs lie within the main sequence with outliers close to the hot-sub dwarfs and white dwarfs as well as a sample in the horizontal and giant branches. It also provides an indication for {\tess} Cycles 1 -- 4 of which TOIs have been identified using SPOC data. The majority of those with SPOC data lie in the main sequence with the exception of a handful of outliers. This aligns with the strategy of selecting SPOC targets.

\subsection{{\tess}-SPOC FFI sensitivity to Transit Detections}
Since detecting transiting exoplanets is the main science goal of the {\tess} mission and each SPOC 2-min light curve is scanned for TCEs, it is worth exploring the sensitivity of the sample light curves to transit detections. Therefore, we explored what would be the radius detection limit from two detected {\tess} transits you could expect to detect for each of the {\tess}-SPOC FFI targets. Hereafter, we refer to this as the two transit radius detection limit. Note that this is an estimate designed to study the sample, rather than a strict limit for individual targets. Furthermore, we do not consider the number of transits which might be observed in total by {\tess}, therefore, we provide only upper-limits on the two transit detectable planet radius. In other words, smaller planets may be detected with many more transits.

To do this we extracted the precision of each {\tess}-SPOC FFI target taken from the SPOC light curve header where provided. The photometric precision metric within the SPOC light curves is based on the Kepler pipeline \citep{jenkins2010transiting} and is called the combined differential photometric precision (CDPP). It is defined as the root mean square of the photometric noise on transit timescales and is used by the SPOC when searching for periodic transit searches. For our purposes, we use the CDPP of each target, the Multiple Event Statistic (MES) threshold of 7.1 from SPOC as a measure of signal-to-noise used to claim a detection and assume a transit duration of 2.7~hours which has been taken as the median transit duration of all the known TOIs. While MES is not identical to signal-to-noise, it is similar according to \citet{jenkins2002impact, twicken2018kepler}. We use this to be consistent with the SPOC pipeline, however, other studies such as \citet{sullivan2015transiting, kunimoto2022predicting} have used the  threshold in signal-to-noise of 7.3. This is put together to estimate the two transit radius detection limit $(\rpl_\text{min})$ from two {\tess} transits in Equation~\ref{eq: Rmin}:

\begin{equation}
    \rpl_\text{min} = 10^{-6}\times \rstar\sqrt{\frac{\text{SNR}\sigma}{\sqrt{T_\text{dur}/2}}}
    \label{eq: Rmin}
\end{equation}

where $\text{SNR}$ is the signal-to-noise threshold of $7.1/\sqrt{2}$ to account for a two transit detection, $T_\text{dur}$ is the average TOI transit duration of 2.7~hrs (divided by 2 since we are using the 2-hour CDPP values), \rstar is the stellar radius and $\sigma$ is the CDPP precision of the lightcurve.

In Figure \ref{fig:planet_radius_HRD} we plot individual HRDs color-coded by the two transit radius detection limit to show the spread in the context of the stellar properties for each {\tess}-SPOC FFI Target. Figure \ref{fig:SPOC_min_radius_hist}, then shows the same data of the two transit radius detection limit as a distribution, where the peak of the histogram lies at $\sim$ 6~\Rearth. Both plots are split up according to planetary radius corresponding to our solar system planets as; Earths/Super-Earths in the range 0.5 $<$ \Rearth $\leq$ 2.0, Mini-Neptunes in the range 2.0 $<$ \Rearth $\leq$ 4.0 , Gas Giants in the range 4.0 $<$ \Rearth $\leq$ 27.0 and Non-Planetary Companions in the range 27.0 $<$ \Rearth $\leq$ 100.0. As can be expected the Earth/Super-Earth population is mostly found amongst the lower main sequence population and Mini-Neptunes are found around stars on the whole of the main sequence. For the larger gas giants these tend to be spread slightly on the main sequence but are also found around stars which are evolved on the horizontal and giant branches. In \S \ref{sec:conclusions}, we compare our findings with that of other planet detection estimate results, namely that if \citet{kunimoto2022predicting}.

\section{Non Single Stars/ Binaries}
\label{sec:binary}
Binary stars can be observed from Earth as visual/resolved, spectroscopic, eclipsing and astrometric. These gravitationally bound star systems are considered to be important as they allow the masses of stars to be determined \citep[see][]{soderhjelm1999visual, al2021comparison}. It is expected that up to 50\% of all stars are in binary systems with some even as triple or higher-multiple systems. However, this fraction is higher for hotter stars, can decrease with higher metallicity and also is related to age and distance \citep{tian2018binary}. As a result of this, Gaia provides its own parameters to indicate the potential binary nature of a given star. 

\begin{figure*}
    \centering
    \includegraphics[width = 0.97\textwidth]{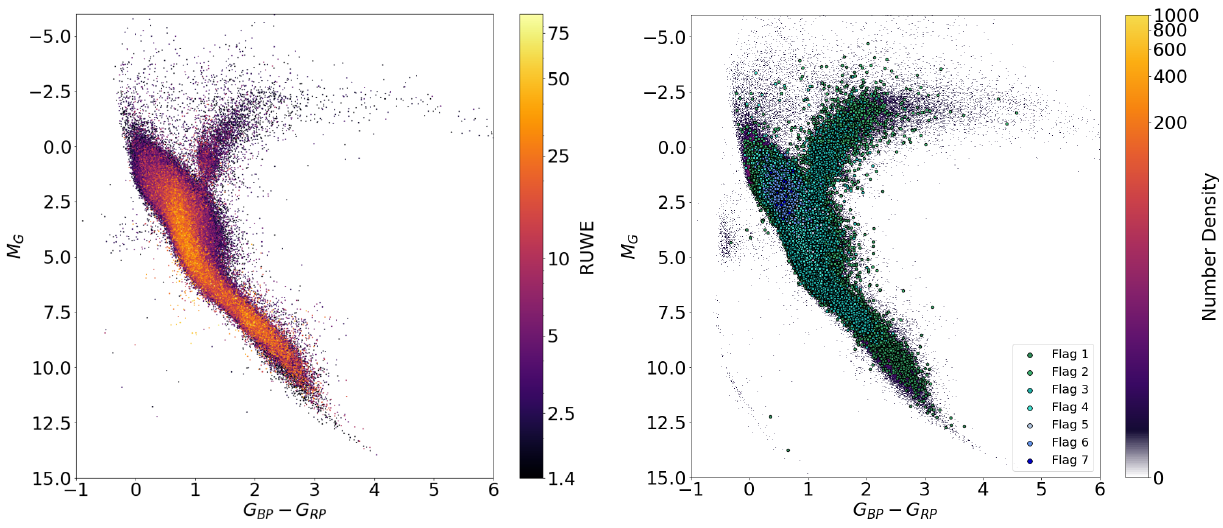}
    \caption{The colour-magnitude diagram of all {\tess}-SPOC FFI Targets from {\tess} Sectors 1 -- 55. The left plot is colour coded according to Gaia RUWE and the right plot has the Gaia non-single star (NSS) flags over plotted. In each plot the higher RUWE and NSS flags are plotted on the top in order to be seen given the large sample size. Full details of what each NSS Flag represents are given in Table \ref{tab:non_single_bits}. }
    \label{fig:binaries_HRD}
\end{figure*}

Firstly, to indicate if any of our stars are members of binary systems we used the Renormalised Unit Weight Error (RUWE) from the Gaia DR3 Catalog \citep[see][for full details]{lindegren2018ruwe}. The RUWE is a goodness-of fit measurement of the single-star model to the targets astrometry which is highly sensitive to the photocentre motion of binaries \citep[see][]{belokurov2020unresolved}. Overall, the RUWE is expected to be around 1.0 for all sources where the single-star model provides a good fit to the astrometric measurements where a RUWE~$\ge$~1.4 could indicate the star is non-single. In total, there are 540,038 {\tess}-SPOC FFI targets with a RUWE~$\ge$~1.4 which could be considered as potential binary systems, which are all plotted in Figure \ref{fig:binaries_HRD}. In Figure \ref{fig:binaries_HRD}, the RUWE value of each {\tess}-SPOC FFI target is shown as the colour of each point on the HRD. Here the more yellow the data point, the higher the RUWE value of the target and the more likely it is in a binary system. Overall, there is a spread of targets with RUWE~$\ge$~1.4 across the whole of the HRD including the main sequence, red giant and horizontal branches and even hot-sub dwarfs and white dwarfs. However, the majority of stars with very high RUWE values are those along the main sequence. This follows what we know about binary systems where approximately half of F, G and K stars are expected to be in a binary system \citep{raghavan2010survey, moe2017mind}. 

\begin{table}
    \centering
        \caption{The number of each {\tess}-SPOC FFI targets which fall into each of the non-single star Flags along with the descriptor for each. The final column shows the percentage of the {\tess}-SPOC FFI Target Sample of 2,890,583 sources.  }
    \begin{tabular}{lccc}
    \hline
         NSS  & Descriptor & Number of & \% of {\tess}- \\
         Flag &  &  {\tess}-SPOC & SPOC FFI Target \\
              &  & Targets & Sample \\
         \hline
         1 & astrometric binary (AB) & 107271 & 3.71\\
         2 & spectroscopic binary (SB) & 56626 &  1.96\\
         3 & AB \& SB & 30642 & 1.06\\
         4 & eclipsing binary (EB) & 1488 & 0.05\\
         5 & AB \& EB& 54 & 0.002\\
         6 & EB \& SB& 296 & 0.01\\
         7 & AB, SB \& EB& 14 & 0.0005\\
         \hline
    \end{tabular}
    \label{tab:non_single_bits}
\end{table}

In addition to the RUWE value, Gaia also provides a non-single star (NSS) flag as an indication of the binary nature of a target. This flag indicates the target has been identified as a NSS by the Gaia Data Processing and Analysis Consortium (DPAC). As such, the observations have been put through tests of various binary orbit models. The NSS solutions are then kept when significant and fit with an acceptable quality. For those where binarity was detected in several instruments a combination fit is considered to improve the precision of the orbital parameters. Therefore, the NSS flag is organised in three main Flags which informs on the nature of the NSS model. They are as follows: Flag 1 is an astrometric binary, Flag 2 is a spectroscopic binary and Flag 4 is an eclipsing binary. Some of the NSS flags represent models which are combinations of these three, for example Flag 5 would represent the combinations of Flag 1 and Flag 4 which informs us that there is a astrometric and eclipsing binary solution (full details of all the flags are in Table \ref{tab:non_single_bits}).

In total, 196,391 {\tess}-SPOC FFI targets have a NSS flag with the combinations within these shown in Table \ref{tab:non_single_bits}. The majority of targets have NSS flag 1 which indicates an astrometric binary. In Figure \ref{fig:binaries_HRD}, each of the individual NSS flags are over-plotted onto the HRD to show the spread. It can be seen that the majority lie on the main sequence and red giant/horizontal branches with only two close to the white dwarf region. Furthermore, the higher NSS flags seem to correlate around the top of the main sequence in the region of the O, B and A stars. This seems reasonable as more than 70\% of OBA stars exist in binary systems \citep{sana2012binary}. This could be a result of the average number of companions per OB primary being at least three times higher than that of low-mass stars with 0.5 companions on average \citep{zinnecker2003formation, grellmann2013multiplicity}. 

Overall, both the Gaia binarity parameters have flagged 18.68\% from RUWE and 6.79\% from NSS in the {\tess}-SPOC FFI targets. In total, 157,206 {\tess}-SPOC FFI targets have both a RUWE~$\ge$~1.4 and NSS flag which equates to 5.43\% of the sample. Therefore, a total of 579,223 {\tess}-SPOC FFI targets have either a RUWE~$\ge$~1.4 or a NSS which equates to 20.04\% of the sample. This is a considerable fraction of the target sample, therefore, for future detailed analysis of these potential binary targets the Gaia information could be used to infer more details on the binary nature. This could be useful when considering searching for planets around these stars and/or looking at variability. 

\section{Isolating the SPOC Main Sequence}
\label{sec:SPOC_ms}
\begin{figure*}
    \centering
    \includegraphics[width = 0.97\textwidth]{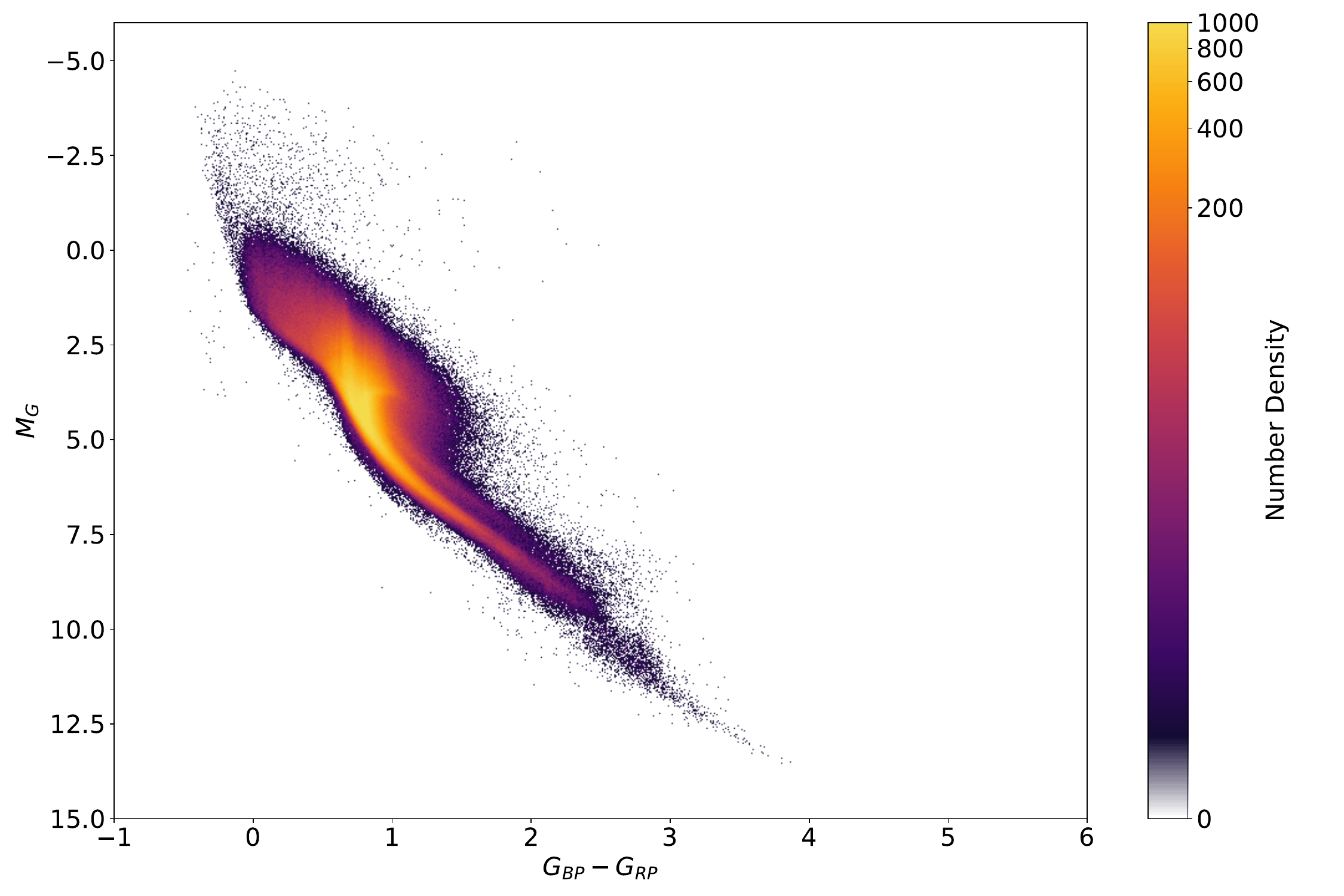}
    \caption{The Gaia colour-magnitude diagram of the {\tess}-SPOC FFI main sequence Target sample. These targets were isolated by filtering on $\log{g} > $ 3.5 which resulted in a final main sequence {\tess}-SPOC FFI Target sample of $\sim$~2.3 million. }
    \label{fig:SPOC_ms}
\end{figure*}

One of the main goals of this paper was to create a comprehensive main sequence {\tess}-SPOC FFI Target sample which can be utilised in other studies. In the creation of the {\tess}-SPOC FFI Target sample \citep{caldwell2020spoctargets}, targets were added up to 160,000 for each sector using the criteria of surface gravity $\log{g} >$~3.5. This added dwarfs and sub-giants to the sample which matches our criteria for selecting the main sequence. Therefore, we apply this criteria to the whole of the {\tess}-SPOC FFI Target sample which leaves us with 2,319,308 main sequence targets. The full HRD of this sample can be seen in Figure \ref{fig:SPOC_ms} where the main sequence is clearly identified and the horizontal and giant branches are no longer seen. The properties for each target, including the Gaia DR3 properties, are listed in Table \ref{tab:SPOC_ms_sample} which is also available as a machine readable table online. 

\begin{table*}
    \centering
    \caption{A small sub sample of the main sequence {\tess}-SPOC FFI Target sample. In total we show a selection of 13 columns out of the 21 available for each target. Amongst these we include the TIC and Gaia DR3 IDs and also the estimated detectable two transit radius detection limit for each target as the last column. This table is available in its entirety in a machine-readable format at CDS where a small sub sample is shown here for guidance regarding its form and content. }
    \begin{tabular}{llccccccccccc}
        \hline
         TIC ID & Gaia DR3 source id  & RA  & DEC  &  parallax  &  g mag  & M$_{\rm{G}}$  & G$_{\rm{BP}}$ - G$_{\rm{RP}}$  & T$_{\rm{eff}}$  & $\log{g}$  &  RUWE  &  NSS  &  {two-transit} radius  \\
                &   & (deg) & (deg)  & (mas)  & & & & (K) & (cgs) & & & (\Rearth) \\  
        \hline
        990573 & 3803801020782388736 & 165.46 & -0.66 & 1.64 & 13.40 & 4.48 & 0.91 & 5569 & 4.23 & 1.09 & 0 & 9.86 \\ 
        25176115 & 1991484152278341632 & 350.97 & 51.21 & 6.00 & 2.65 & 6.54 & 1.38 & 4512 & 4.48 & 3.21 & 1 & 5.43 \\
        66457200 & 477598718345633920 & 73.69 & 61.76 & 2.66 & 9.22 & 1.35 & 0.29 & 8121 & 3.88 & 0.94 & 0 & 7.84 \\
        284468412 & 4559345936350050432 & 253.24 & 17.29 & 5.57 & 10.85 & 4.59 & 0.93 & 5496 & 4.29 & 1.39 & 2 & 4.43 \\ 
        420831735 & 5929340070299722368 & 251.61 & -55.57 & 1.33& 10.81 & 1.44 & 0.63 & 7037 & 3.46 & 0.81 & 0 & 27.6 \\    \vdots &  \vdots &  \vdots &  \vdots &  \vdots&  \vdots &  \vdots &  \vdots &  \vdots &  \vdots &  \vdots &  \vdots &  \vdots \\
        \hline
    \end{tabular}
    \label{tab:SPOC_ms_sample}
\end{table*}

\section{Discussion \& Conclusions}
\label{sec:conclusions}
In this paper we have explored the properties of the {\tess}-SPOC FFI Target sample by utilising the Gaia DR3 catalogue. From this we have identified a well defined main sequence {\tess}-SPOC FFI Target sample which is available to the community for future studies (see Table \ref{tab:SPOC_ms_sample}). The aim behind this, is to standardise a sample which can be used for large surveys which are based on the {\tess}-SPOC FFI data which is publicly available.

Firstly, we created our final {\tess}-SPOC FFI target sample by cross-matching the {\tess}-SPOC FFI Target sample with Gaia DR3 using the Gaia DR2 source identifier. Any {\tess}-SPOC FFI targets which were missed out were matched using a cone search with Gaia on the right ascension and declination to form a more complete sample, see \S \ref{sec:final_sample} for the full details. The comparisons between both the Gaia sample and the {\tess}-SPOC FFI Target sample with regards to absolute magnitude show a stronger spread amongst main sequence targets for SPOC, see Figure \ref{fig:Gaia_hist}. The mean absolute Gaia magnitude for the Gaia sample is 2.6 and for the {\tess}-SPOC FFI Target sample is 3.9 (see Figure \ref{fig:Gaia_hist}), which falls inline with the selection criteria for SPOC and is unsurprising. 

For each of our {\tess}-SPOC FFI targets, we used the CDPP from the SPOC FFI light curves to estimate the two transit radius detection limit of an orbiting exoplanet. In order to do this, we also had to make a few assumptions: (i) the transit duration was set to 2.7~hours (median transit duration of all known TOIs) and (ii) the signal-to-noise threshold was set to $7.1/\sqrt{2}$. In \citet{kunimoto2022predicting}, they use simulations of {\tess} data along with injection and recovery tests to determine the number of detectable planets around 8.5~million AFGKM stars from the {\tess} Candidate Target List (CTL: v8.01). Their stellar sample was taken from the CTL keeping only targets which were classified as dwarfs and removing all giants. This is different to the {\tess}-SPOC FFI Target sample which, as seen from Figure \ref{fig:SPOC_HRD}, has targets from main sequence dwarfs to red giants and white dwarfs. In total, over seven years of {\tess} observations \citet{kunimoto2022predicting} predict a planet yield of 12,519 $\pm$ 678, with 8426 $\pm$ 525 planets detectable from the primary mission and first extended mission (i.e. years 1 - 4). It is important to note, they have an upper limit of 16~\Rearth for AFGK stars and 4~\Rearth for M stars within their injection and recovery tests. Overall, they find that G-type stars are the most common hosts and half of the {\tess} planet detections consist of gas giants with radii greater than 8~\Rearth. 

From our {\tess}-SPOC FFI main sequence Target sample, 41\% of targets are G-type stars which is in agreement with our median $T_{\rm{eff}}$ of 5,933~K, therefore, the majority of our two transit radii detection limits are from host stars which are G-type. Furthermore, we also find that 38\% of our two transit radii detection limits are giants with $R_p >$ 8~\Rearth. We also have a peak in our distribution of detectable planetary radii which falls at $\sim$~6~\Rearth. While it is important to note we are not producing a yield estimate, it is reassuring that similarities are found amongst our two transit radii detection limits. We do not perform any injection and recovery tests as it is beyond the scope of this paper. Furthermore, \citet{kunimoto2022predicting} used a model planet distribution which is a function of planet size, orbital period, and host star spectral type to predict their yield. In this study we only present the detectability, but do not consider how it varies with increasing numbers of transits, where this would lead to smaller planet radii detections. Furthermore, when calculating our two transit radii we assume uncorrelated Gaussian noise. However, the {\tess} SPOC pipeline goes to considerable effort to decorrelate the data before any transit detections, therefore, the noise within the light curves is generally not uncorrelated Gaussian noise and does not strictly obey square-root statistics. There is no simple solution to correct for this, but we would just like to make a note here for the reader.

With our cross-matched sample we explored the Gaia properties for the {\tess}-SPOC FFI Target sample, see Figure \ref{fig:SPOC_prop_hist}. The median effective temperature of the sample is 5,933~K, this lies in the range of G-type stars and is in agreement with our stellar spectral type distribution. Further to this, the median stellar radius is 1.36~\Rsun which is a typical radius for an FGK star. There also appears to be a second bump in radius at $\sim$ 0.5~\Rsun which is indicative of low mass M dwarfs. This can also be seen in distance at $\sim$100~pc and in effective temperature at $\sim$3000~K which all correspond to low mass M dwarf properties. This is not surprising since {\tess} has a redder band-pass than Kepler, making it easier to detect exoplanets around low mass stars. Furthermore, the increase in M dwarfs with {\tess} can also be attributed to their emphasis in the selection process for SPOC 2-min and SPOC FFI targets, see \S \ref{sec:SPOC_sample}. With regards to distance, the median was 438~pc with 90\% of stars having a distance of 850~pc or less. Finally, the median $\log{g}$ is 4.25 which centres around what we know for the Sun with $\log{g}$ = 4.43 \citep{gray2021observation}. There is also a second peak in the $\log{g}$ distribution at 4.5 suggesting a group of high surface gravity stars, however, there is also a small peak at $\sim$0.3~\Rsun. Therefore, it is likely the peaks in $\log{g}$ of 4.5, T$_{\rm{eff}}$ of $\sim$ 3,000~K and radius of $\sim$0.3~\Rsun correspond to M dwarfs which forms part of the SPOC selection process.

Overall, this paper presents the first overview of the {\tess}-SPOC FFI Target sample, utilising data from Gaia DR3. We have produced the first HRD of the {\tess}-SPOC FFI Target sample showing a wide distribution of targets from the main sequence, red giant branch, white dwarfs and hot-sub dwarfs. Furthermore, we have produced a main sequence {\tess}-SPOC FFI Target sample which is publicly available for the community to use in further studies. Given the success of the SPOC pipeline and the planet yield expected from TESS going forward, we hope this will encourage a standardised target list to be used when appropriate. Furthermore, we have also estimated the two transit radius detection limit for each target using {\tess}-SPOC FFI data to inform future planet searches. This may help when developing search strategies for new transiting exoplanets in the future. 

\section*{Acknowledgements}
We include data in this paper collected by the {\tess} mission, where funding for the {\tess} mission is provided by the NASA Explorer Program. This work presents results from the European Space Agency (ESA) space mission Gaia. Gaia data are being processed by the Gaia Data Processing and Analysis Consortium (DPAC). Funding for the DPAC is provided by national institutions, in particular the institutions participating in the Gaia MultiLateral Agreement (MLA). The Gaia mission website is \url{https://www.cosmos.esa.int/gaia}. The Gaia archive website is \url{https://archives.esac.esa.int/gaia}.

This research was funded in whole or in part by the UKRI, (Grants ST/X001121/1, EP/X027562/1). The authors would like to thank the referee for their comments and suggestions which helped to improve the paper. For the purpose of open access, the author has applied a Creative Commons Attribution (CC BY) licence (where permitted by UKRI, ‘Open Government Licence’ or ‘Creative Commons Attribution No-derivatives (CC BY-ND) licence’ may be stated instead) to any Author Accepted Manuscript version arising from this submission.

\section*{Data Availability}

All {\tess} SPOC data are available from the NASA MAST portal and Gaia DR2 and DR3 data is available from the Gaia Archive. We have included the final {\tess} {\tess}-SPOC FFI main sequence target sample as supplementary material online for availability to the community.



\bibliographystyle{mnras}
\bibliography{Gaia_SPOC} 

\begin{thebibliography}{}
\makeatletter
\relax
\def\mn@urlcharsother{\let\do\@makeother \do\$\do\&\do\#\do\^\do\_\do\%\do\~}
\def\mn@doi{\begingroup\mn@urlcharsother \@ifnextchar [ {\mn@doi@}
  {\mn@doi@[]}}
\def\mn@doi@[#1]#2{\def\@tempa{#1}\ifx\@tempa\@empty \href
  {http://dx.doi.org/#2} {doi:#2}\else \href {http://dx.doi.org/#2} {#1}\fi
  \endgroup}
\def\mn@eprint#1#2{\mn@eprint@#1:#2::\@nil}
\def\mn@eprint@arXiv#1{\href {http://arxiv.org/abs/#1} {{\tt arXiv:#1}}}
\def\mn@eprint@dblp#1{\href {http://dblp.uni-trier.de/rec/bibtex/#1.xml}
  {dblp:#1}}
\def\mn@eprint@#1:#2:#3:#4\@nil{\def\@tempa {#1}\def\@tempb {#2}\def\@tempc
  {#3}\ifx \@tempc \@empty \let \@tempc \@tempb \let \@tempb \@tempa \fi \ifx
  \@tempb \@empty \def\@tempb {arXiv}\fi \@ifundefined
  {mn@eprint@\@tempb}{\@tempb:\@tempc}{\expandafter \expandafter \csname
  mn@eprint@\@tempb\endcsname \expandafter{\@tempc}}}

\bibitem[\protect\citeauthoryear{Al-Wardat, Hussein, Al-Naimiy  \&
  Barstow}{Al-Wardat et~al.}{2021}]{al2021comparison}
Al-Wardat M.~A.,  Hussein A.~M.,  Al-Naimiy H.~M.,   Barstow M.~A.,  2021,
  Publications of the Astronomical Society of Australia, 38, e002

\bibitem[\protect\citeauthoryear{Babusiaux et~al.,}{Babusiaux
  et~al.}{2018}]{babusiaux2018gaia}
Babusiaux C.,  et~al., 2018, \aap, 616, A10

\bibitem[\protect\citeauthoryear{Bailer-Jones et~al.,}{Bailer-Jones
  et~al.}{2013}]{bailer2013gaia}
Bailer-Jones C.,  et~al., 2013, \aap, 559, A74

\bibitem[\protect\citeauthoryear{Battley, Pollacco  \& Armstrong}{Battley
  et~al.}{2020}]{battley2020search}
Battley M.~P.,  Pollacco D.,   Armstrong D.~J.,  2020, \mnras, 496, 1197

\bibitem[\protect\citeauthoryear{Belokurov et~al.,}{Belokurov
  et~al.}{2020}]{belokurov2020unresolved}
Belokurov V.,  et~al., 2020, \mnras, 496, 1922

\bibitem[\protect\citeauthoryear{Brown et~al.,}{Brown
  et~al.}{2018}]{brown2018gaiadr2}
Brown A.,  et~al., 2018, \aap, 616, A1

\bibitem[\protect\citeauthoryear{Brown et~al.,}{Brown
  et~al.}{2021}]{brown2021gaia}
Brown A.~G.,  et~al., 2021, \aap, 649, A1

\bibitem[\protect\citeauthoryear{Bryant, Bayliss  \& Van~Eylen}{Bryant
  et~al.}{2023}]{bryant2023occurrence}
Bryant E.~M.,  Bayliss D.,   Van~Eylen V.,  2023, \mnras, 521, 3663

\bibitem[\protect\citeauthoryear{Caldwell et~al.,}{Caldwell
  et~al.}{2020}]{caldwell2020spoctargets}
Caldwell D.~A.,  et~al., 2020, Research Notes of the AAS, 4, 201

\bibitem[\protect\citeauthoryear{Davenport, Mendoza  \& Hawley}{Davenport
  et~al.}{2020}]{davenport202010}
Davenport J.~R.,  Mendoza G.~T.,   Hawley S.~L.,  2020, \aj, 160, 36

\bibitem[\protect\citeauthoryear{Doyle, Bagnulo, Ramsay, Doyle  \&
  Hakala}{Doyle et~al.}{2022}]{doyle2022puzzling}
Doyle L.,  Bagnulo S.,  Ramsay G.,  Doyle J.~G.,   Hakala P.,  2022, Monthly
  Notices of the Royal Astronomical Society, 512, 979

\bibitem[\protect\citeauthoryear{Eisner et~al.,}{Eisner
  et~al.}{2021}]{eisner2021planet}
Eisner N.~L.,  et~al., 2021, \mnras, 501, 4669

\bibitem[\protect\citeauthoryear{Gandolfi et~al.,}{Gandolfi
  et~al.}{2018}]{gandolfi2018tess}
Gandolfi D.,  et~al., 2018, \aap, 619, L10

\bibitem[\protect\citeauthoryear{Gilbert et~al.,}{Gilbert
  et~al.}{2020}]{gilbert2020first}
Gilbert E.~A.,  et~al., 2020, \aj, 160, 116

\bibitem[\protect\citeauthoryear{Gill et~al.,}{Gill
  et~al.}{2020}]{gill2020ngts}
Gill S.,  et~al., 2020, \apjl, 898, L11

\bibitem[\protect\citeauthoryear{Gray}{Gray}{2021}]{gray2021observation}
Gray D.~F.,  2021, The observation and analysis of stellar photospheres.
Cambridge university press

\bibitem[\protect\citeauthoryear{Grellmann, Preibisch, Ratzka, Kraus, Helminiak
   \& Zinnecker}{Grellmann et~al.}{2013}]{grellmann2013multiplicity}
Grellmann R.,  Preibisch T.,  Ratzka T.,  Kraus S.,  Helminiak K.,   Zinnecker
  H.,  2013, \aap, 550, A82

\bibitem[\protect\citeauthoryear{Guerrero et~al.,}{Guerrero
  et~al.}{2021}]{guerrero2021tois}
Guerrero N.~M.,  et~al., 2021, \apjs, 254, 39

\bibitem[\protect\citeauthoryear{Huang et~al.,}{Huang
  et~al.}{2018}]{huang2018tess}
Huang C.~X.,  et~al., 2018, \apjl, 868, L39

\bibitem[\protect\citeauthoryear{IJspeert, Tkachenko, Johnston, Garcia,
  De~Ridder, Van~Reeth  \& Aerts}{IJspeert et~al.}{2021}]{ijspeert2021all}
IJspeert L.~W.,  Tkachenko A.,  Johnston C.,  Garcia S.,  De~Ridder J.,
  Van~Reeth T.,   Aerts C.,  2021, \aap, 652, A120

\bibitem[\protect\citeauthoryear{Jenkins}{Jenkins}{2002}]{jenkins2002impact}
Jenkins J.~M.,  2002, \apj, 575, 493

\bibitem[\protect\citeauthoryear{Jenkins et~al.,}{Jenkins
  et~al.}{2010}]{jenkins2010transiting}
Jenkins J.~M.,  et~al., 2010, in Software and Cyberinfrastructure for
  Astronomy. pp 140--150

\bibitem[\protect\citeauthoryear{Jenkins et~al.,}{Jenkins
  et~al.}{2016}]{jenkins2016tessspoc}
Jenkins J.~M.,  et~al., 2016, in Software and Cyberinfrastructure for Astronomy
  IV. pp 1232--1251

\bibitem[\protect\citeauthoryear{Kunimoto, Winn, Ricker  \&
  Vanderspek}{Kunimoto et~al.}{2022}]{kunimoto2022predicting}
Kunimoto M.,  Winn J.,  Ricker G.~R.,   Vanderspek R.~K.,  2022, \aj, 163, 290

\bibitem[\protect\citeauthoryear{Leleu et~al.,}{Leleu
  et~al.}{2021}]{leleu2021six}
Leleu A.,  et~al., 2021, \aap, 649, A26

\bibitem[\protect\citeauthoryear{Lendl et~al.,}{Lendl
  et~al.}{2020}]{lendl2020toi}
Lendl M.,  et~al., 2020, \mnras, 492, 1761

\bibitem[\protect\citeauthoryear{Lindegren et~al.}{Lindegren
  et~al.}{2018}]{lindegren2018ruwe}
Lindegren L.,  et~al., 2018, Gaia Technical Note: GAIA-C3-TN-LU-LL-124-01

\bibitem[\protect\citeauthoryear{Mann et~al.,}{Mann
  et~al.}{2022}]{mann2022tess}
Mann A.~W.,  et~al., 2022, \aj, 163, 156

\bibitem[\protect\citeauthoryear{Moe \& Di~Stefano}{Moe \&
  Di~Stefano}{2017}]{moe2017mind}
Moe M.,  Di~Stefano R.,  2017, \apjs, 230, 15

\bibitem[\protect\citeauthoryear{Newton et~al.,}{Newton
  et~al.}{2019}]{newton2019tess}
Newton E.~R.,  et~al., 2019, \apjl, 880, L17

\bibitem[\protect\citeauthoryear{Oddo et~al.,}{Oddo
  et~al.}{2023}]{oddo2023characterization}
Oddo D.,  et~al., 2023, \aj, 165, 134

\bibitem[\protect\citeauthoryear{Pr{\v{s}}a et~al.,}{Pr{\v{s}}a
  et~al.}{2022}]{prvsa2022tess}
Pr{\v{s}}a A.,  et~al., 2022, \apjs, 258, 16

\bibitem[\protect\citeauthoryear{Prusti et~al.,}{Prusti
  et~al.}{2016}]{prusti2016gaia}
Prusti T.,  et~al., 2016, Astronomy \& astrophysics, 595, A1

\bibitem[\protect\citeauthoryear{Raghavan et~al.,}{Raghavan
  et~al.}{2010}]{raghavan2010survey}
Raghavan D.,  et~al., 2010, \apjs, 190, 1

\bibitem[\protect\citeauthoryear{Ricker et~al.,}{Ricker
  et~al.}{2015}]{ricker2015tess}
Ricker G.~R.,  et~al., 2015, Journal of Astronomical Telescopes, Instruments,
  and Systems, 1, 014003

\bibitem[\protect\citeauthoryear{Sana et~al.,}{Sana
  et~al.}{2012}]{sana2012binary}
Sana H.,  et~al., 2012, Science, 337, 444

\bibitem[\protect\citeauthoryear{Smith, Morris, Jenkins, Bryson, Caldwell  \&
  Girouard}{Smith et~al.}{2016}]{smith2016finding}
Smith J.~C.,  Morris R.~L.,  Jenkins J.~M.,  Bryson S.~T.,  Caldwell D.~A.,
  Girouard F.~R.,  2016, \pasp, 128, 124501

\bibitem[\protect\citeauthoryear{S{\"o}derhjelm}{S{\"o}derhjelm}{1999}]{soderhjelm1999visual}
S{\"o}derhjelm S.,  1999, \aap, 341, 121

\bibitem[\protect\citeauthoryear{Stassun et~al.,}{Stassun
  et~al.}{2018}]{stassun2018tess}
Stassun K.~G.,  et~al., 2018, \aj, 156, 102

\bibitem[\protect\citeauthoryear{Stassun et~al.,}{Stassun
  et~al.}{2019}]{stassun2019tic}
Stassun K.~G.,  et~al., 2019, \aj, 158, 138

\bibitem[\protect\citeauthoryear{Sullivan et~al.,}{Sullivan
  et~al.}{2015}]{sullivan2015transiting}
Sullivan P.~W.,  et~al., 2015, \apj, 809, 77

\bibitem[\protect\citeauthoryear{Taylor}{Taylor}{2005}]{taylor2005topcat}
Taylor M.~B.,  2005, in Astronomical data analysis software and systems XIV.
  p.~29

\bibitem[\protect\citeauthoryear{Tian et~al.,}{Tian
  et~al.}{2018}]{tian2018binary}
Tian Z.-J.,  et~al., 2018, Research in Astronomy and Astrophysics, 18, 052

\bibitem[\protect\citeauthoryear{Twicken et~al.,}{Twicken
  et~al.}{2018}]{twicken2018kepler}
Twicken J.~D.,  et~al., 2018, \pasp, 130, 064502

\bibitem[\protect\citeauthoryear{Vallenari et~al.,}{Vallenari
  et~al.}{2023}]{vallenari2023gaiadr3}
Vallenari A.,  et~al., 2023, \aap, 674, A1

\bibitem[\protect\citeauthoryear{Yee et~al.,}{Yee et~al.}{2022}]{yee2022tess}
Yee S.~W.,  et~al., 2022, \aj, 164, 70

\bibitem[\protect\citeauthoryear{Zinnecker}{Zinnecker}{2003}]{zinnecker2003formation}
Zinnecker H.,  2003, in Symposium-International Astronomical Union. pp 80--90

\makeatother
\end{thebibliography}







\bsp	
\label{lastpage}
\end{document}